\documentclass[aps,prd,groupedaddress,twocolumn, 10pt, superscriptaddress, notitlepage, longbibliography]{revtex4-2}

\usepackage{lipsum}
\usepackage{amsmath}
\usepackage{amssymb}
\usepackage{amsthm}
\usepackage{bbm}
\usepackage{graphicx}
\usepackage{color}
\usepackage{upgreek}
\usepackage[linkcolor = blue, citecolor = blue, urlcolor = blue, colorlinks = true]{hyperref}
\usepackage{amssymb}
\usepackage{bm}
\usepackage[capitalise]{cleveref}
\usepackage{textcomp}
\usepackage{hyperref}
\usepackage[usenames,dvipsnames]{xcolor}
\usepackage[normalem]{ulem}
\usepackage{wasysym}

\newcommand\diff{\mathrm{d}}
\renewcommand{\vec}[1]{\mathbf{#1}}
\renewcommand{\imath}[0]{\mathsf{i}}

\hypersetup{colorlinks=true, linkcolor=blue!50!black, urlcolor=blue!50!black, citecolor=blue!50!black}

\definecolor{ABpurple}{RGB}{128, 0, 128}
\definecolor{ABred}{RGB}{255, 0, 0}
\definecolor{ABgreen}{RGB}{0, 255, 0}
\definecolor{ABbrown}{RGB}{128, 64, 0}
\definecolor{ABblue}{RGB}{0, 0, 255}

\newcommand{\inlinemaketitle}{{\let\newpage\relax\maketitle}}
\allowdisplaybreaks

\begin{document}

\title{Microswimmers near corrugated, periodic surfaces}
\author{Christina Kurzthaler}
\email{ck24@princeton.edu}
\affiliation{Department of Mechanical and Aerospace Engineering, Princeton University, Princeton, New Jersey 08544, USA}
\author{Howard A. Stone}
\email{hastone@princeton.edu}
\affiliation{Department of Mechanical and Aerospace Engineering, Princeton University, Princeton, New Jersey 08544, USA}
\begin{abstract}
We explore hydrodynamic interactions between microswimmers and corrugated, or rough, surfaces, as found often in biological systems and microfluidic devices. Using the Lorentz reciprocal theorem for viscous flows we derive exact expressions for the roughness-induced velocities up to first order in the surface-height fluctuations and provide solutions for the translational and angular velocities valid for arbitrary surface shapes. We apply our theoretical predictions to elucidate the impact of a periodic, wavy surface on the velocities of microswimmers modeled in terms of a superposition of Stokes singularities. Our findings, valid in the framework of a far-field analysis, show that the roughness-induced velocities vary non-monotonically with the wavelength of the surface.  For wavelengths comparable to the swimmer-surface distance a pusher can experience a repulsive contribution due to the reflection of flow fields at the edge of a surface cavity, which decreases the overall attraction to the wall.
\end{abstract}
\maketitle

\section{Introduction}
Biological environments offer a plethora of different geometries and confining surfaces, ranging from elastic and soft boundaries to rough, structured topographies, which impact transport processes in their nearby surroundings~\cite{Hoefling:2013,Persat:2015, Bechinger:2016, Berne:2018,Bhattacharjee:2019, Daddi:2019}. Interactions between particles and surfaces composed of different materials can be of, amongst others, chemical~\cite{Uspal:2019}, electrical~\cite{Poortinga:2002,Humphries:2017}, thermal~\cite{Kroy:2016}, steric~\cite{Lushi:2017}, or hydrodynamic origin~\cite{Leal:2007, Berke:2008}. Unraveling the individual contributions is of utmost importance for our understanding of microbiological phenomena and future progress in micro- and nanotechnological applications.

To optimize survival strategies many microorganisms, including bacteria~\cite{Berg:1972,Berg:2008}, sperm~\cite{Woolley:2003,Riedel:2005,Friedrich:2008}, protozoa~\cite{Machemer:1972}, and algae~\cite{Merchant:2007}, self-propel by using cellular appendages, such as flagella and cilia. In a similar spirit, due to their potential for inspiring novel drug-delivery and bioremediation tools, artificial active agents have been synthesized and rely on various concepts, including self-phoresis~\cite{Howse:2007,Buttinoni:2012,Kurzthaler:2018} and biomimetic propulsion mechanisms~\cite{Dreyfus:2005,Ghosh:2009}.
These out-of-equilibrium systems display intricate interactions with surfaces that differ significantly from those of their passive counterparts. While it is well-known that by time-reversibility of Stokes flow a sphere sedimenting near a vertical, plane wall maintains a constant distance to it~\cite{Oneill:1967,Goldman:1967}, flow fields generated by spherical microswimmers can induce attraction or reorientation of their swimming direction~\cite{Zoettl:2016}. These effects are determined by the details of the propulsion mechanism, such as pusher-~\cite{Drescher:2011} or puller-type~\cite{Drescher:2010} swimming strokes and surface-slip due to cilia~\cite{Drescher:2010} or phoresis~\cite{Uspal:2015}, and the pitch angle at which the agents approach the surface~\cite{Zoettl:2016,Shen:2018}. Thus, fluid-mediated interactions in combination with steric effects have been shown to drive accumulation of self-propelled agents near planar, smooth surfaces~\cite{Fauci:1995, Berke:2008, Smith:2009,Drescher:2011,Spagnolie:2012}.

In addition, hydrodynamic couplings with surfaces can strongly modify the dynamical behavior of flagellated microorganisms, such as \emph{Escherichia coli}~\cite{Diluzio:2005,Lauga:2006}, \emph{Vibrio cholerae}~\cite{Utada:2014}, and sperm~\cite{Woolley:2003, Friedrich:2008,Elgeti:2010}, which display circular swimming patterns near a solid wall. Bacteria are also sensitive to the slip length of the boundary, which can induce circular trajectories along the opposite sense of rotation compared to a planar wall~\cite{diLeonardo:2011}, and a change of surface slip can thereby randomize or direct the motion of nearby bacteria~\cite{Lemelle:2013,Hu:2015}.
Far-field predictions~\cite{Spagnolie:2015}, simulations of the squirmer model~\cite{Kuron:2019}, and experiments with artificial microrods~\cite{Takagi:2014} have shown that active agents can get trapped or scattered by spherical obstacles depending on the geometric features of the obstacles and the propulsion mechanism. A recent study on sperm motion in channels with sharp corners and curved walls have related their detachment from the channel surface at the corners to the orientation of the flagellum beating pattern~\cite{Rode:2019}.

At a larger scale, the presence of random heterogeneities, comparable in size to a bacterial body, distributed on smooth surfaces amplify near-surface dynamics of ~\emph{E.~coli} cells~\cite{Frangipane:2019,Makarchuk:2019}. For example, obstacles can enhance bacterial transport at intermediate obstacle densities by disrupting their circular motion~\cite{Makarchuk:2019}, whereas large densities of surface heterogeneities decrease residence times of bacteria at the surface due to less accessible space~\cite{Frangipane:2019}. Moreover, Brownian dynamics simulations predict that the dispersion of active agents is further affected by the radius of the circular motion compared to the obstacle size~\cite{Chepizhko:2019}.

Near-surface motion entails various unusual features at the microscopic level, where details of particle-surface interactions are important, and at the mesoscopic scale, where the dynamical behavior of these active agents is affected by noise intrinsic to biological systems. In the present study we focus on the microscopic scale and elucidate the hydrodynamic couplings of microswimmers with nearby corrugated surfaces. We provide an analytical framework to calculate velocities induced by the presence of a surface with arbitrary shape up to first order in the surface-height fluctuations. We then apply our theoretical predictions to study far-field hydrodynamics of microswimmers modeled in terms of a multipole expansion.

\section{Model}
\begin{figure}[htp]
\centering
\includegraphics[width=\linewidth]{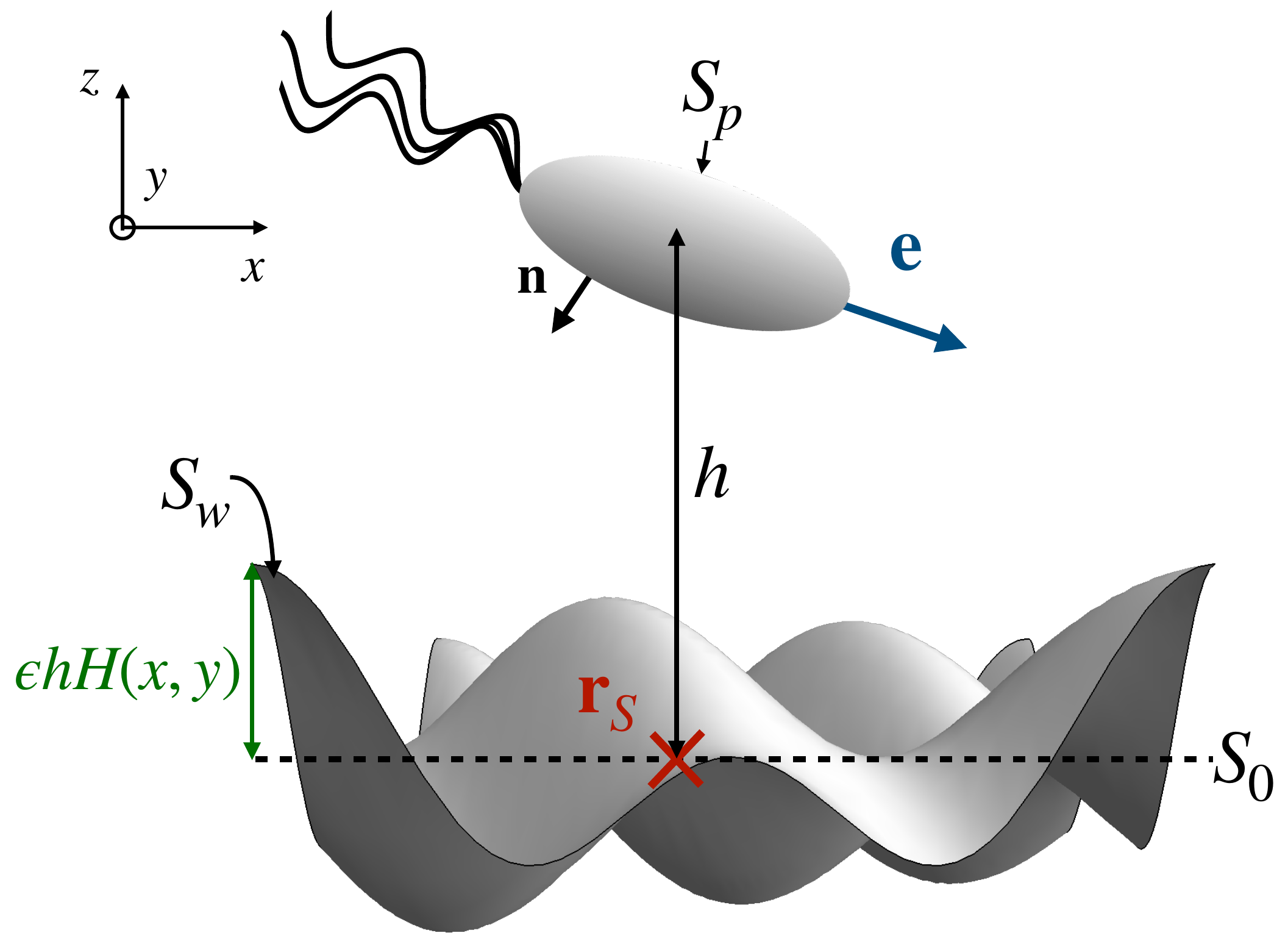}
\caption{Model set-up for the motion of a microswimmer with surface~$S_p$ and swimming direction~$\vec{e}$. It is located at position $\vec{r}_S=[x_S,y_S]^T$ and distance $h$ near a corrugated wall~$S_w$ with shape $H(x,y)$. Furthermore, $\epsilon h$ denotes the amplitude of the surface-height fluctuations and $S_0$ corresponds to the planar, reference surface at $z=0$. \label{fig:sketch}}
\end{figure}

We consider the hydrodynamic interactions of a microswimmer with a nearby corrugated surface, $S_w$, see Fig.~\ref{fig:sketch}.
The active agent is suspended in an incompressible, low-Reynolds-number flow. The quasi-steady fluid velocity $\mathbf{u}(\mathbf{r})$ and pressure fields $p(\mathbf{r})$ are described by the continuity and Stokes equations,
\begin{align}
\nabla\cdot\vec{u} = 0 \quad \text{ and }\quad \mu \nabla^2 \vec{u} &= \nabla p, \label{eq:stokes}
\end{align}
with associated stress field $\boldsymbol{\sigma} = -p\mathbb{I} + \mu\left(\nabla \vec{u}+\nabla \vec{u}^T\right)$ and fluid viscosity $\mu$.

Biological microorganisms and artificial active agents display a large variety of different swimming mechanisms that rely on, e.g., non-reciprocal deformations of flagella, surface distortions generated by cilia, or diffusiophoretic mechanisms. The fluid velocity on the surface of these microswimmers can be decomposed into a translational velocity $\vec{U}$, flow induced by the body rotation at angular velocity $\boldsymbol{\Omega}$, and a disturbance velocity field, $\vec{u}_S$, which can vary locally over the particle surface,~$S_p$ ~\cite{Stone:1996}.
In the laboratory frame the boundary conditions of a microswimmer near a rough wall are
\begin{subequations}
\begin{align}
\vec{u}&= \vec{u}_S+\vec{U}+\vec{r}\times\boldsymbol{\Omega} & \hspace{-3cm} &\text{ on } S_p,\label{eq:bc_sp}\\
\vec{u}&=\vec{0} & & \text{ on } S_w \text{ and } S_\infty, \label{eq:bc_sw}
\end{align}
\end{subequations}
where $S_\infty$ denotes the bounding surface at infinity. The translational and angular velocities, $\vec{U}=\vec{U}(\vec{r}_S(t),h(t))$ and $\boldsymbol{\Omega}=\boldsymbol{\Omega}(\vec{r}_S(t),h(t))$, depend on the instantaneous configuration of the microswimmer relative to the underlying textured surface, $\vec{r}_S= [x_S,y_S]^T$, and its distance to the wall $h$. In principle, the particle surface can change over time, $S_p=S_p(t)$. Subsequently, we suppress the time dependence.

Furthermore, the microswimmer experiences zero hydrodynamic force and torque,
\begin{subequations}
\begin{align}
\vec{F}_H &= \int_{S_p} \, \vec{n}\cdot\boldsymbol{\sigma} \ \diff S = \vec{0} \label{eq:zero_force}\\ 
\vec{L}_H &= \int_{S_p} \, \vec{r}\times \left(\vec{n}\cdot\boldsymbol{\sigma}\right) \ \diff S = \vec{0}. \label{eq:zero_torque}
\end{align}
\end{subequations}
In principle, the velocities of the microswimmer can be obtained by solving the Stokes and continuity equations [Eq.~\eqref{eq:stokes}] with boundary conditions [Eqs.~\eqref{eq:bc_sp}-\eqref{eq:bc_sw}] and satisfying the force- and torque-free conditions [Eq.~\eqref{eq:zero_force}-\eqref{eq:zero_torque}]. Here, we circumvent calculating the full fluid flow by applying the Lorentz reciprocal theorem for viscous flows~\cite{Leal:2007,Masoud:2019}. We provide analytical expressions for the roughness-induced velocities up to first order in the surface-height fluctuations by following a similar procedure as in Ref.~\cite{Kurzthaler:2020}.

\section{Roughness-induced velocities}
We describe the corrugated surface in terms of a dimensionless shape function $H(x,y)$ and consider small height fluctuations compared to the particle-surface distance $h$. The surface assumes the form $z=\epsilon h H(x,y)$, where we have introduced a small dimensionless parameter, $\epsilon\ll 1$.
Thus, we can expand the velocity field in~$\epsilon$:
\begin{align}
\mathbf{u} = \mathbf{u}^{(0)}+\epsilon\mathbf{u}^{(1)} +\mathcal{O}(\epsilon^2),
\end{align}
where $\vec{u}^{(0)}$ denotes the velocity field generated by a microswimmer near a plane, smooth surface and $\vec{u}^{(1)}$ encodes the roughness-induced velocity field.
Consequently, also the translational and rotational velocities of the microswimmer assume the forms,
\begin{subequations}
\begin{align}
\vec{U}&=\vec{U}^{(0)}+\epsilon\vec{U}^{(1)}+\mathcal{O}(\epsilon^2), \\
\boldsymbol{\Omega}&=\boldsymbol{\Omega}^{(0)}+\epsilon\boldsymbol{\Omega}^{(1)}+\mathcal{O}(\epsilon^2).
\end{align}
\end{subequations}
Using the method of domain perturbations~\cite{Kamrin:2010}, we expand the no-slip boundary condition at the corrugated surface~$S_w$ in terms of a Taylor expansion about $z=0$,
\begin{align}
\begin{split}
\mathbf{u}&(x,y,\epsilon hH(x,y))=\\
&\mathbf{u}^{(0)}(x,y,0)+\epsilon h H(x,y)\frac{\partial \mathbf{u}^{(0)}}{\partial z}\Bigr|_{z=0}\!+\epsilon\mathbf{u}^{(1)}(x,y,0) +\mathcal{O}(\epsilon^2).
\end{split}
\end{align}
Collecting orders in $\epsilon$, we obtain the boundary conditions for the zeroth- and first-order problems,
\begin{align}
\vec{u}^{(0)} &= \vec{0} \quad \text{ and } \quad
\vec{u}^{(1)} = -hH(x,y)\frac{\partial \mathbf{u}^{(0)}}{\partial z}\Bigr|_{z=0} \quad \text{ on } S_0. \label{eq:bc_1}
\end{align}
The Taylor expansion of the boundary condition provides the replacement of the no-slip boundary condition at the corrugated surface, $S_w$, by an effective slip velocity at the plane, smooth surface~$S_0$. Therefore, the relevant boundaries of our problem constitute the surface of the sphere~$S_p$, the plane reference surface~$S_0$, and the bounding surface at infinity~$S_\infty$.
The derived boundary conditions allow us to fully describe the zeroth- and first-order problems.

\subsection{Zeroth-order problem: planar wall}
The zeroth-order problem with flow field $\mathbf{u}^{(0)}$ corresponds to a microswimmer moving near a plane wall. It satisfies the Stokes and continuity equations, $\nabla\cdot\boldsymbol{\sigma}^{(0)}=\vec{0}$ and $\nabla\cdot\mathbf{u}^{(0)}=0$, respectively, with boundary conditions
\begin{subequations}
\begin{align}
\vec{u}^{(0)}&=\vec{u}_S+\vec{U}^{(0)}+\vec{r}\times\boldsymbol{\Omega}^{(0)} & \hspace{-1.5cm} &\text{ on } S_p,\\
\vec{u}^{(0)}&= \vec{0} & &\text{ on } S_0 \text{ and } S_\infty,
\end{align}
\end{subequations}
where $\vec{u}_S$ is specified. Analytical progress has been made for the paradigmatic squirmer model, which was introduced by Lighthill to model the motion of nearly spherical, deformable microswimmers~\cite{Lighthill:1952}. Exact solutions for the velocity field $\vec{u}^{(0)}$ are available for a spherical squirmer of radius $a$ near a planar wall in terms of a bispherical coordinate representation valid for arbitrary particle-surface distances $h/a$~\cite{Shaik:2017} and the lubrication approximation, $h/a\lesssim 1$~\cite{Ishikawa:2006}.
In addition, the far-field flows generated by self-propelled particles can be described in terms of a multipole expansion in Stokes singularities~\cite{Berke:2008, Spagnolie:2012}. We refer to Sec.~\ref{sec:far_field} for further details.

\subsection{First-order problem: surface roughness}
The first-order correction to the fluid flow due to the underlying textured surface, $\mathbf{u}^{(1)}$, obeys the Stokes and continuity equations, $\nabla\cdot\boldsymbol{\sigma}^{(1)}=\vec{0}$ and $\nabla\cdot\mathbf{u}^{(1)}=0$, with boundary conditions:
\begin{subequations}
\begin{align}
\vec{u}^{(1)}&=\vec{U}^{(1)}+\vec{r}\times\boldsymbol{\Omega}^{(1)} & &\text{ on } S_p, \label{eq:bc_1_all}\\
\vec{u}^{(1)}&= -hH(x,y)\frac{\partial \mathbf{u}^{(0)}}{\partial z}\Bigr|_{z=0}& &\text{ on } S_0,\\
\vec{u}^{(1)}&=\vec{0} & &\text{ on } S_\infty. \label{eq:bc_3_all}
\end{align}
\end{subequations}
Here, the effective slip velocity at the planar surface,~$S_0$, involves the surface shape $H(x,y)$ and the zeroth-order flow field, $\vec{u}^{(0)}$.

We develop an analytic theory for the translational and rotational velocities of a microswimmer near a corrugated wall by exploiting the Lorentz reciprocal theorem for viscous flows~\cite{Leal:2007, Masoud:2019}. The reciprocal theorem relates Stokes flows in a given domain that obey different sets of boundary conditions. Therefore, we introduce as an auxiliary problem the flow due to an externally driven, translating and rotating particle of the same shape as the swimmer near a plane wall. The corresponding velocity field $\hat{\vec{u}}$ satisfies the no-slip boundary conditions $\hat{\vec{u}}=\hat{\vec{U}}+\mathbf{r}\times\hat{\boldsymbol{\Omega}}$ on $S_p$ and $\hat{\vec{u}}=\vec{0}$ on $S_0$ and $S_\infty$. The applied force, $\hat{\vec{F}}$, and torque, $\hat{\vec{L}}$, are balanced by the hydrodynamic force, $\hat{\vec{F}}_H$, and torque, $\hat{\vec{L}}_H$, on the particle.
Then the reciprocal theorem relates our main problem with velocity field $\vec{u}^{(1)}$ and stress tensor $\boldsymbol\sigma^{(1)}$ to the auxiliary problem with $\hat{\vec{u}}$, $\hat{\boldsymbol{\sigma}}$ by
\begin{align}
\int_{S_p \cup\, S_0\, \cup\, S_\infty} \vec{n}\cdot \boldsymbol{\sigma}^{(1)} \cdot \hat{\vec{u}} \ \diff S &= \int_{S_p \cup\, S_0\, \cup\, S_\infty} \vec{n}\cdot \hat{\boldsymbol{\sigma}} \cdot \vec{u}^{(1)} \ \diff S, \label{eq:lorentz}
\end{align}
where the integrals are taken over all surfaces, i.e., $S_p$, $S_0$, and $S_\infty$.
Here, we employ the notation that the normal vector $\vec{n}$ is directed away from the corresponding surface into the surrounding fluid (see Fig.~\ref{fig:sketch}). Since the main problem is force- and torque-free [Eq.~\eqref{eq:zero_force}-\eqref{eq:zero_torque}] and given the boundary conditions on $\hat{\vec{u}}$, the left-hand side of Eq.~\eqref{eq:lorentz} vanishes. Utilizing the boundary conditions of the main [Eqs.~\eqref{eq:bc_1_all}-\eqref{eq:bc_3_all}] and auxiliary problems, the reciprocal relation [Eq.~\eqref{eq:lorentz}] simplifies to
\begin{align}
\begin{split}
\hat{\mathbf{F}}_H\cdot\mathbf{U}^{(1)} &+ \hat{\mathbf{L}}_H\cdot\boldsymbol{\Omega}^{(1)}=\\
&\int_{S_0} \, \vec{n}\cdot \hat{\boldsymbol{\sigma}}\cdot \left(hH(x,y)\frac{\partial \mathbf{u}^{(0)}}{\partial z}\Bigr|_{z=0}\right)\ \diff S.
\end{split}\label{eq:lorentz_U1}
\end{align}
This relation is exact for small surface height fluctuations and valid for arbitrary microswimmer and surface shapes. Thus, for any known zeroth-order problem $\vec{u}^{(0)}$ and appropriate auxiliary problem $\hat{\vec{u}}$ the first-order correction to the swimming velocities due to the corrugated surface shape can be calculated by evaluating the expression derived from Eq.~\eqref{eq:lorentz_U1}.

\section{Far-field hydrodynamics of a microswimmer near a wavy, periodic surface\label{sec:far_field}}
\begin{figure}[htp]
\centering
\includegraphics[width=\linewidth]{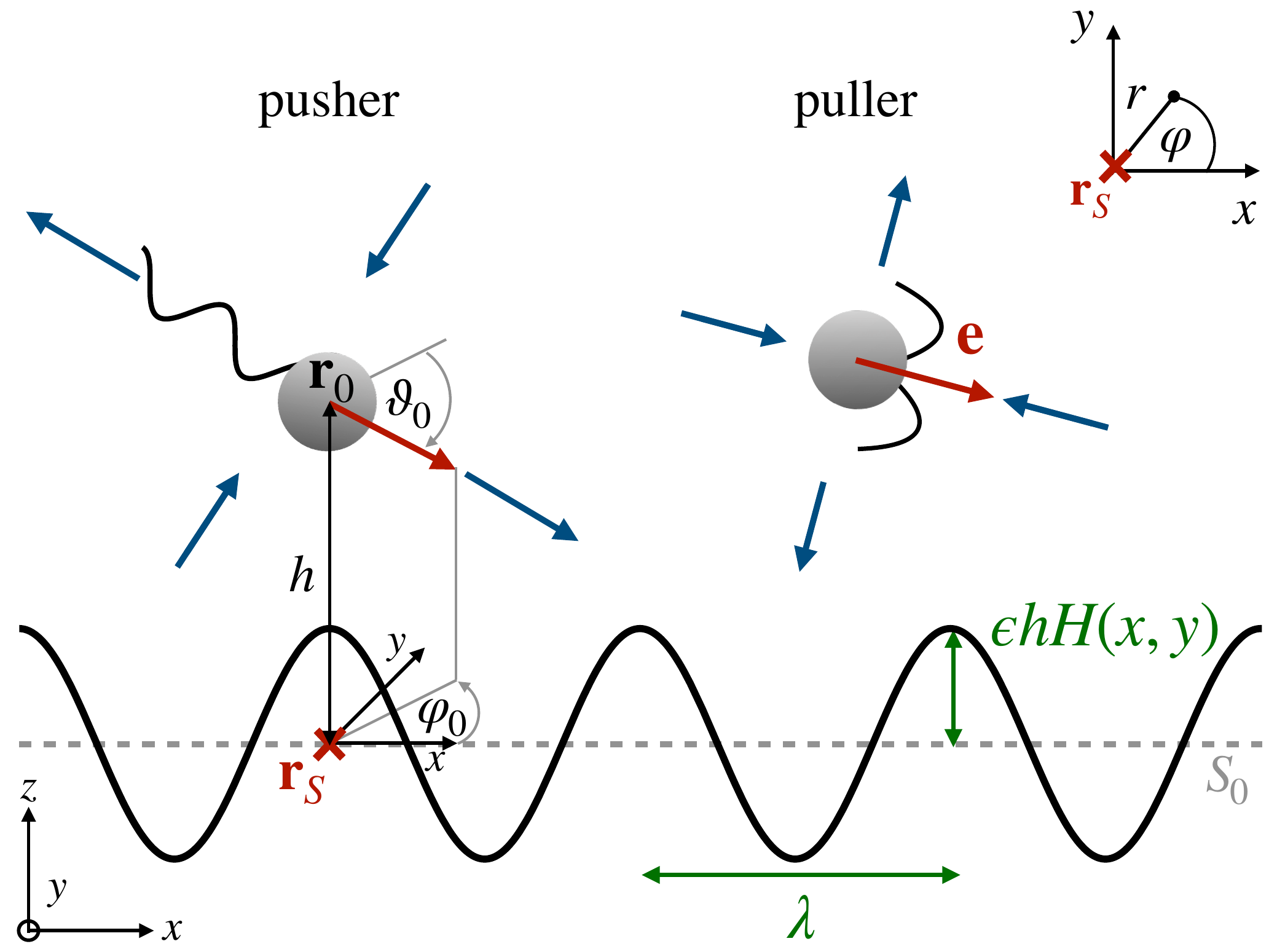}
\caption{Sketch of pusher- and puller-type microswimmers with swimming direction $\vec{e}$, pitch and polar angles, $\vartheta_0$ and $\varphi_0$, near a wavy surface, $H(x,y)= \cos(2\pi x/\lambda)$, with characteristic wavelength~$\lambda$. In particular, the pitch angle measures the angle with respect to a horizontal line parallel to the plane reference surface, $S_0$, and the polar angle, $\varphi_0$, measures the orientation in the $x-y$ plane. The blue arrows indicate the flow direction for the corresponding force dipoles in an unbounded fluid. The local coordinate system with radial distance $r$ and polar angle $\varphi$ is displayed in the upper right corner.
\label{fig:sketch_ff}}
\end{figure}

We apply our theoretical predictions to elucidate roughness-induced velocities of a spherical microswimmer with radius $a$ located at a distance $h/a\gtrsim 1$ away from the surface. We model the flow-field generated by the microswimmer in terms of a far-field description~\cite{Spagnolie:2012}. The microswimmer is located at a position $\vec{r}_0 = [\vec{r}_S,h]^T$ near a corrugated surface and its swimming direction, $\vec{e}= \vec{e}_{\varphi_0}\cos\vartheta_0+\vec{e}_z\sin\vartheta_0$ with $\vec{e}_{\varphi_0}=\vec{e}_x\cos\varphi_0+\vec{e}_y\sin\varphi_0$,
is characterized by the pitch and polar angles, $\vartheta_0$ and $\varphi_0$. The unit vector $\vec{e}_{\varphi_0}$ measures
the swimming direction $\vec{e}$ in a plane parallel to the reference surface $S_0$. The polar angle $\varphi_0$ characterizes the swimmer's orientation in the $x-y$ plane. The angle $\vartheta_0$ is measured from a horizontal line parallel to the plane reference surface, $S_0$, at the height from the swimmer center (see Fig.~\ref{fig:sketch_ff}). In particular, for $\vartheta_0=0$ the swimmer is aligned parallel to $S_0$ and for $\vartheta_0=\pm\pi/2$ it is aligned perpendicular away from or toward $S_0$, respectively.

The flow field generated in an unbounded domain can be decomposed in terms of a multipole expansion~\cite{Spagnolie:2012},
\begin{align}
\vec{u} &= \vec{u}_{FD}+\vec{u}_{SD}+\vec{u}_{FQ}+\vec{u}_{RD}+\mathcal{O}(|\vec{r}-\vec{r}_0|^{-4})
\end{align}
where the terms correspond to contributions of a force dipole (FD), $\vec{u}_{FD}= \alpha_{FD} \mathbf{G}_{FD}(\vec{e}, \vec{e})$,
source dipole (SD), $\vec{u}_{SD} = \alpha_{SD} \vec{G}_{SD}(\vec{e})$, force quadrupole (FQ), $\vec{u}_{FQ} = \alpha_{FQ} \vec{G}_{FQ}(\vec{e},\vec{e},\vec{e})$, and rotlet dipole (RD), $\vec{u}_{RD} = \alpha_{RD}\vec{G}_{RD}(\vec{e},\vec{e})$. We follow the notation of Ref.~\cite{Spagnolie:2012} and introduce the corresponding singularity solutions $\vec{G}_{FD, SD,FQ,RD}$ and singularity strengths $\alpha_{FD, SD,FQ,RD}$ below.
The contributions due to higher-order singularities can be derived from the stokeslet solution at position $\vec{r}_0$ directed along $\vec{e}$,
\begin{align}
\mathbf{G}(\vec{r}, \vec{r}_0;\vec{e}) = \left(\vec{e}+(\vec{e}\cdot\hat{\vec{r}})\hat{\vec{r}}\right)/\hat{r},
\end{align}
with $\hat{r}=|\vec{r}-\vec{r}_0|$ and $\hat{\mathbf{r}}=\left(\vec{r}-\vec{r}_0\right)/\hat{r}$.
The flow field produced by a stokeslet can then be written as $\vec{u}=\alpha \mathbf{G}(\vec{r}, \vec{r}_0;\vec{e}) $, where the singularity strength $\alpha$ is related to the magnitude of the force $F$ and the viscosity $\mu$ via $\alpha = F/(8\pi\mu)$. The flow field of a force dipole with two point forces separated by a distance $\ell$ along the direction $\vec{a}$ can be obtained as
\begin{subequations}
\begin{align}
\vec{u}_{FD} &=\alpha [\mathbf{G}(\vec{r}, \vec{r}_0+\ell\vec{a}/2;\vec{e})\!-\!\mathbf{G}(\vec{r}, \vec{r}_0-\ell\vec{a}/2;\vec{e})] \\
&\simeq\alpha_{FD} (\vec{a}\cdot\nabla_0)\mathbf{G}(\vec{r}, \vec{r}_0;\vec{e}), \label{eq:FD_2}
\end{align}
\end{subequations}
where Eq.~\eqref{eq:FD_2} remains valid for small $\ell$ and we have introduced the force dipole strength $\alpha_{FD}=\alpha\ell$. The gradient $\nabla_0$ acts on the singularity position $\vec{r}_0$. This allows us to introduce the force dipole singularity solution
\begin{align}
\mathbf{G}_{FD}(\vec{e},\vec{a})\!\equiv\!\mathbf{G}_{FD}(\vec{r},\vec{r}_0;\vec{e},\vec{a})\!=\!(\vec{a}\cdot\nabla_0)\mathbf{G}(\vec{r}, \vec{r}_0;\vec{e}).
\end{align}
In particular, the velocity field induced by a force dipole oriented along $\vec{e}$ can be expressed as
\begin{align}
\vec{u}_{FD}
&=\alpha_{FD} \bigl(\mathbf{G}_{FD}(\vec{e}_{\varphi_0},\vec{e}_{\varphi_0}) \cos^2\vartheta_0 + \mathbf{G}_{FD}(\vec{e}_z,\vec{e}_z) \sin^2\vartheta_0 \notag\\
&\phantom{=\alpha} \ \ +\mathbf{G}_{SS}(\vec{e}_{\varphi_0},\vec{e}_z) \sin(2\vartheta_0)\bigr),
\end{align}
 where $\vec{G}_{SS}(\vec{a},\vec{b})=\frac{1}{2}\left(\vec{G}_{FD}(\vec{a},\vec{b})+\vec{G}_{FD}(\vec{b},\vec{a})\right)$
denotes the symmetric part of the stokeslet, also referred to as a stresslet.
Similar relations hold for the other oriented higher-order singularities. In particular, the force quadrupole singularity solution can be obtained from the force dipole solution via
\begin{align}
\begin{split}
    \vec{G}_{FQ}(\vec{e,\vec{a},\vec{b}}) &\equiv \vec{G}_{FQ}(\vec{r}_0,\vec{r};\vec{e,\vec{a},\vec{b}})\\
    &= \left(\vec{b}\cdot\nabla_0\right)\mathbf{G}_{FD}(\vec{r}_0,\vec{r};\vec{e},\vec{a}).
    \end{split}
\end{align}
The source dipole singularity solution can be expressed in terms of the Stokeslet solution via
\begin{align}
\vec{G}_{SD}(\vec{e}) &\equiv  \vec{G}_{SD}(\vec{r}, \vec{r}_0;\vec{e})= -\frac{1}{2}\nabla_0^2\vec{G}(\vec{r},\vec{r}_0;\vec{e}).
\end{align}
Finally, the singularity solution for the rotlet dipole is
\begin{align}
    \vec{G}_{RD}(\vec{e},\vec{c})\equiv \vec{G}_{RD}(\vec{r},\vec{r}_0;\vec{e},\vec{c}) = \left(\vec{c}\cdot\nabla_0\right)\vec{G}_R(\vec{r},\vec{r}_0; \vec{e}),
\end{align}
which depends on the singularity solution of a rotlet $\vec{G}_R(\vec{r},\vec{r}_0; \vec{e})=\left[\vec{G}_{FD}(\vec{b},\vec{a})-\vec{G}_{FD}(\vec{a},\vec{b})\right]/2$ with unit vectors $\vec{a}$ and $\vec{b}$ obeying $\vec{a}\times\vec{b}=\vec{e}$.
We note that the leading-order flow field, in the absence of a wall, is generated by the force dipole and decays as $\hat{r}^{-2}$. The next higher-order-singularities (SD, FQ, and RD) decay as $\hat{r}^{-3}$.

The associated singularity strengths, $\alpha_{FD}, \alpha_{SD},\alpha_{FQ},\alpha_{RD}$, depend on the details of the swimming mechanisms. In particular, the dipole strength $\alpha_{FD}$  has dimensions of [velocity$\times$length$^2$] whereas the higher-order singularity strengths have dimensions of [velocity$\times$length$^3$].
The force dipole strength allows distinguishing between particles that produce extensile flow fields (pushers, $\alpha_{FD}>0$) and contractile flow fields (pullers, $\alpha_{FD}<0$). The far-field hydrodynamics induced by the finite size of the swimming object can be described in terms of the source dipole. Typically, the sign of the source dipole strength for ciliated microswimmers is positive $\alpha_{SD}>0$, whereas for flagellated organisms it is negative $\alpha_{SD}<0$, which indicates repulsion from a wall for a swimmer with orientation away from the wall and an attraction for a swimmer with orientation towards the wall. The flow fields generated by a swimming body with fore-aft symmetry can be modeled by a force quadrupole; in particular, $\alpha_{FQ}>0$ corresponds to swimmers with long flagella compared to the body size and vice versa for $\alpha_{FQ}<0$. The rotlet dipole can be used to describe, e.g., the flow field produced by the rotation of the flagellum and the cell body, which can induce clockwise ($\alpha_{RD}>0$) or counter-clockwise ($\alpha_{RD}<0$) swimming motion along surfaces. For more details we refer to Ref.~\cite{Spagnolie:2012}.

\subsection{Smooth, planar wall}
The velocity field induced by a spherical microswimmer located at $\vec{r}_0$ near a smooth, planar wall can be decomposed into the velocity field generated in an unbounded domain, $\vec{u}$, and the disturbance velocity field, $\vec{u}^*$, due to the nearby wall using the image method adapted from electrostatics~\cite{Blake:1974, Spagnolie:2012},
\begin{align}
\vec{u}^{(0)} &= \vec{u}+\vec{u}^*.
\end{align}
The wall-induced velocity field $\vec{u}^*$ depends on the position of the image singularity, $\vec{r}^*_0=\vec{r}_0-2h\vec{e}_z$. Its complete dependence on the fundamental solutions of Stokes flow has been provided earlier~\cite{Spagnolie:2012}.

Faxen's law predicts the translational and angular velocities of the microswimmer at position $\vec{r}_0$ induced by the wall~\cite{Kim:2005},
\begin{align}
\vec{U}^* &= \vec{u}^*(\vec{r}_0) +\mathcal{O}(a^2\nabla^2\vec{u}^*|_{\vec{r}_0}), \\
\boldsymbol{\Omega}^* &= \frac{1}{2}\nabla\times\vec{u}^*(\vec{r}_0) + \mathcal{O}(a^2\nabla^2(\nabla\times \vec{u}^*)|_{\vec{r}_0}). \label{eq:omega_fd}
\end{align}
Then the translational and rotational velocities of the microswimmer near a plane wall evaluate to
$\vec{U}^{(0)} =\vec{U}_\text{free} + \vec{U}^*$ and $\boldsymbol{\Omega}^{(0)} =\boldsymbol{\Omega}^*$,
where we have introduced the translational velocity of a microswimmer in an unbounded domain $\vec{U}_\text{free}=U\vec{e}$ and $\vec{U}^*, \boldsymbol{\Omega}^*$ include contributions from the swimmer-wall interactions. In particular, for a force dipole these evaluate to~\cite{Berke:2008,Spagnolie:2012}
\begin{subequations}
\begin{align}
\vec{U}^*_{FD}&\!=\!\frac{3\alpha_{FD}}{16h^2} \left(1\!-\!3\cos (2\vartheta_0)\right)\vec{e}_z\!+\!\frac{3\alpha_{FD}}{8h^2}\sin(2\vartheta_0)\vec{e}_{\varphi_0}, \\
\boldsymbol{\Omega}^*_{FD} &\!=\!\frac{3\alpha_{FD}}{16h^3}\sin(2\vartheta_0)\vec{e}_{\varphi_0}^\perp,
\end{align}
\end{subequations}
with $\vec{e}^\perp_{\varphi_0} = \vec{e}_z\times\vec{e}_{\varphi_0}$. The wall-induced velocities for the higher-order singularities are provided in Appendix~\ref{appendix:vel}.

\subsection{Roughness-induced velocities}
To obtain the contribution due to the wall roughness,  $\vec{U}^{(1)}$ and $\boldsymbol{\Omega}^{(1)}$ (see Eq.~\eqref{eq:bc_1_all}), we apply the reciprocal relation [Eq.~\eqref{eq:lorentz_U1}]. We use as auxiliary problems the cases of a point force and point torque parallel to the $x-y$ plane and $z$-directions, respectively. Furthermore, we introduce a local, cylindrical coordinate system $(r,\varphi,z)$ where $r = \sqrt{(x-x_S)^2+(y-y_S)^2}$ denotes the distance measured from the particle position in the $x-y-$plane relative to a point on the surface and $\varphi$ is the polar angle (see inset of Fig.~\ref{fig:sketch_ff}). The gradient of the flow field induced by the microswimmer at the wall then assumes the form
\begin{align}
\frac{\partial \vec{u}^{(0)}}{\partial z}\Bigr|_{z=0} &= \frac{\partial u^{(0)}_r}{\partial z}\Bigr|_{z=0}\vec{e}_r + \frac{\partial u^{(0)}_\varphi}{\partial z}\Bigr|_{z=0}\vec{e}_\varphi,
\end{align}
with unit vectors $\vec{e}_r=\vec{e}_x\cos\varphi+\vec{e}_y\sin\varphi$ and $\vec{e}_\varphi=-\vec{e}_x\sin\varphi+\vec{e}_y\cos\varphi$. The derivative of the $z-$component at the no-slip surface vanishes by continuity, $\left[\partial u^{(0)}_z/\partial z\right]_{z=0}=0$.
Further, we rescale the coordinates by the particle-surface distance, $z = hZ$ and $r=hR$, the velocities by $\vec{u}^{(0)} = \alpha_{FD} \vec{U}^{(0)}/h^2$ and the stresses by
\begin{align}
\hat{\boldsymbol{\sigma}}^{i} &=\frac{ \hat{F}^{i}}{h^2}\hat{\boldsymbol{\Sigma}}^{U,i} \qquad \text{ and }  \qquad \hat{\boldsymbol{\sigma}}^{i} =\frac{ \hat{L}^{i}}{h^3}\hat{\boldsymbol{\Sigma}}^{\Omega,i},
\end{align}
respectively. Here, $\hat{\vec{F}}^{i}=\hat{F}^{i}\vec{e}_i$ and $\hat{\vec{L}}^{i}=\hat{L}^{i}\vec{e}_i$ denote the point force and torque along the $i-$direction ($i=x,y,z$), which are balanced by the hydrodynamic force and torque: $\hat{\vec{F}}^{i}=-\hat{\vec{F}}_H^i$ and $\hat{\vec{L}}^{i}=-\hat{\vec{L}}_H^i$.
\begin{figure*}[htp]
\centering
\includegraphics[width = 0.75\linewidth]{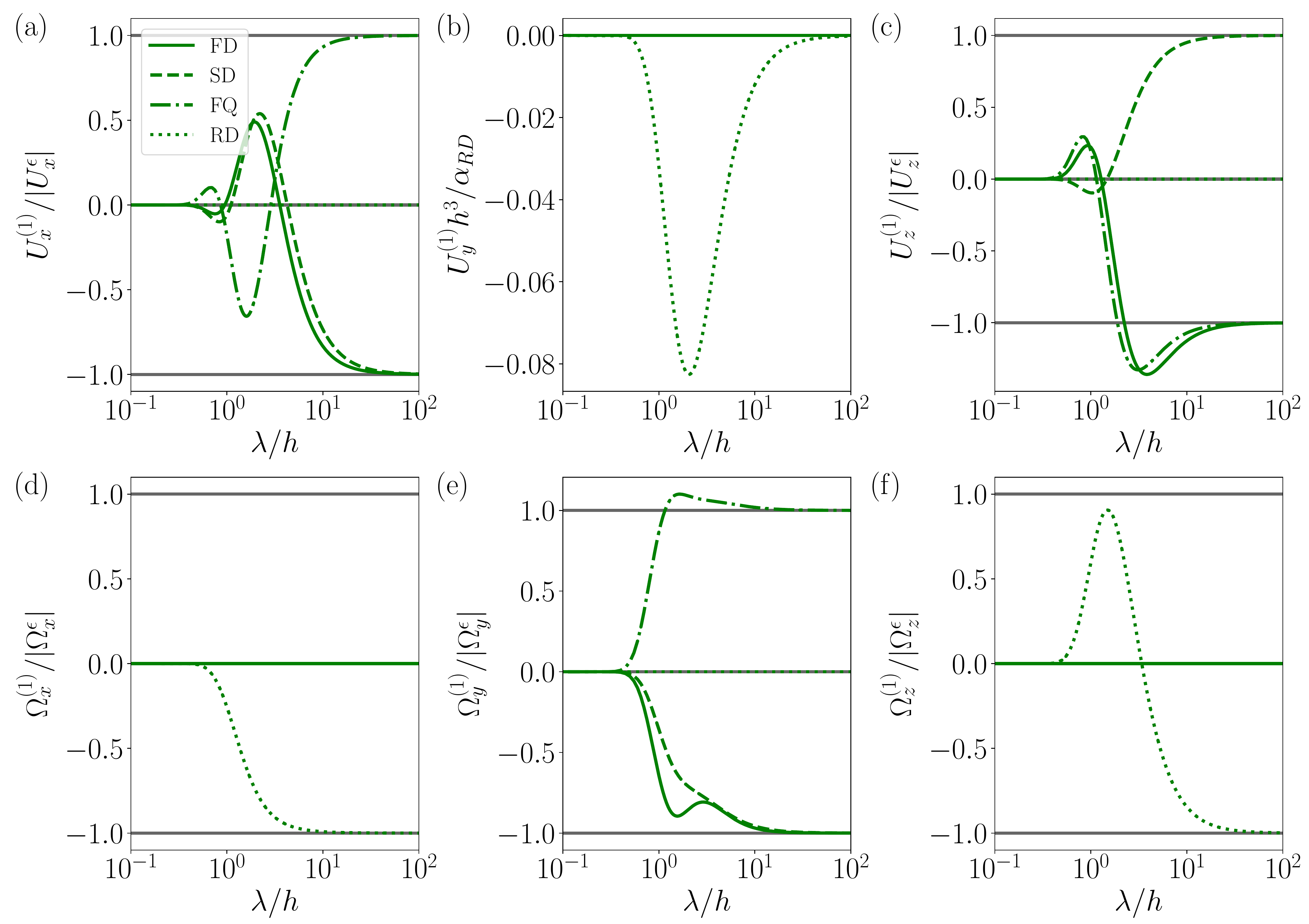}
\caption{Roughness-induced velocities of a force dipole (FD), source dipole (SD), force quadrupole (FQ), and rotlet dipole (RD) near a wavy surface of shape $H(x,y)=\cos(2\pi x/\lambda)$. The components of the first-order velocities, $\vec{U}^{(1)}$ (a-c) and $\boldsymbol{\Omega}^{(1)}$ (d-f), are shown with respect to the wavelength $\lambda/h$ for a microswimmer located at a distance $h/a=2$, at position $x_S/\lambda = 0$, and with pitch angle $\vartheta_0=-\pi/8$ and $\varphi_0=0$.
The velocities are normalized by the velocities induced by a wall shifted closer to the particle by $\epsilon h$, $\vec{U}^\epsilon = d_U \vec{U}^*$ and $\boldsymbol{\Omega}^\epsilon = d_\Omega \boldsymbol{\Omega}^*$. Here, $d_U=2$ ($d_U=3$) and $d_\Omega=3$ ($d_\Omega=4$) for the force dipole (all other higher-order singularities).
\label{fig:velocities_x0}}
\end{figure*}
\nopagebreak
Then, the components of the first-order translational and rotational velocities are obtained by evaluating the surface integrals,
\begin{subequations}
\begin{align}
\begin{split}
U^{(1)}_i &= -\frac{\alpha_{FD}}{ h^2}\int_0^\infty\, \int_0^{2\pi} \, \vec{n}\cdot\hat{\boldsymbol{\Sigma}}^{U,i}\cdot\\
&\qquad
 \left(H(\vec{r}_S; R, \varphi)\frac{\partial \vec{U}^{(0)}}{\partial Z}\Bigr|_{Z=0} \right) \! R \ \diff \varphi \ \diff R,
\end{split}
\end{align}
\begin{align}
\begin{split}
\Omega^{(1)}_i &= -\frac{\alpha_{FD}}{ h^3}\int_0^\infty\, \int_0^{2\pi} \, \vec{n}\cdot \hat{\boldsymbol{\Sigma}}^{\Omega,i} \cdot\\
&\qquad\left(H(\vec{r}_S; R, \varphi)\frac{\partial \vec{U}^{(0)}}{\partial Z}\Bigr|_{Z=0} \right) \! R \ \diff \varphi \ \diff R.
\end{split}
\end{align}
\end{subequations}
By exploiting the symmetries of the stresses these equations simplify to
\begin{subequations}
\begin{align}
\begin{split}
U^{(1)}_i &= -\frac{\alpha_{FD}}{ h^2}\int_0^\infty\, \int_0^{2\pi} \, H(\vec{r}_S; R, \varphi)\times\\
& \phantom{-\frac{\alpha}{6\pi h^2}\int_0^\infty\, \int_0^{2\pi}}
\left[\hat{\Sigma}^{U,i}_{ZR}\frac{\partial U^{(0)}_{R}}{\partial Z} \right]_{Z=0} \! R \ \diff \varphi \ \diff R,
\end{split}\label{eq:U}
\end{align}
\begin{align}
\begin{split}
\Omega^{(1)}_i &= -\frac{\alpha_{FD}}{ h^3}\int_0^\infty\, \int_0^{2\pi} \,  H(\vec{r}_S; R, \varphi)\times\\
& \left[\hat{\Sigma}^{\Omega,i}_{ZR} \frac{\partial U^{(0)}_{R}}{\partial Z}+ \hat{\Sigma}^{\Omega,i}_{Z\varphi} \frac{\partial U^{(0)}_{\varphi}}{\partial Z}\right]_{Z=0} \! R \ \diff \varphi \ \diff R.
\end{split}\label{eq:O}
\end{align}
\end{subequations}
These expressions are valid for arbitrary surface shapes $H(x,y)$ and depend on the position $\vec{r}_S$ and distance $h$ of the microswimmer relative to the underlying surface.

\begin{figure*}[tp]
\centering
\includegraphics[width=0.75\linewidth,keepaspectratio]{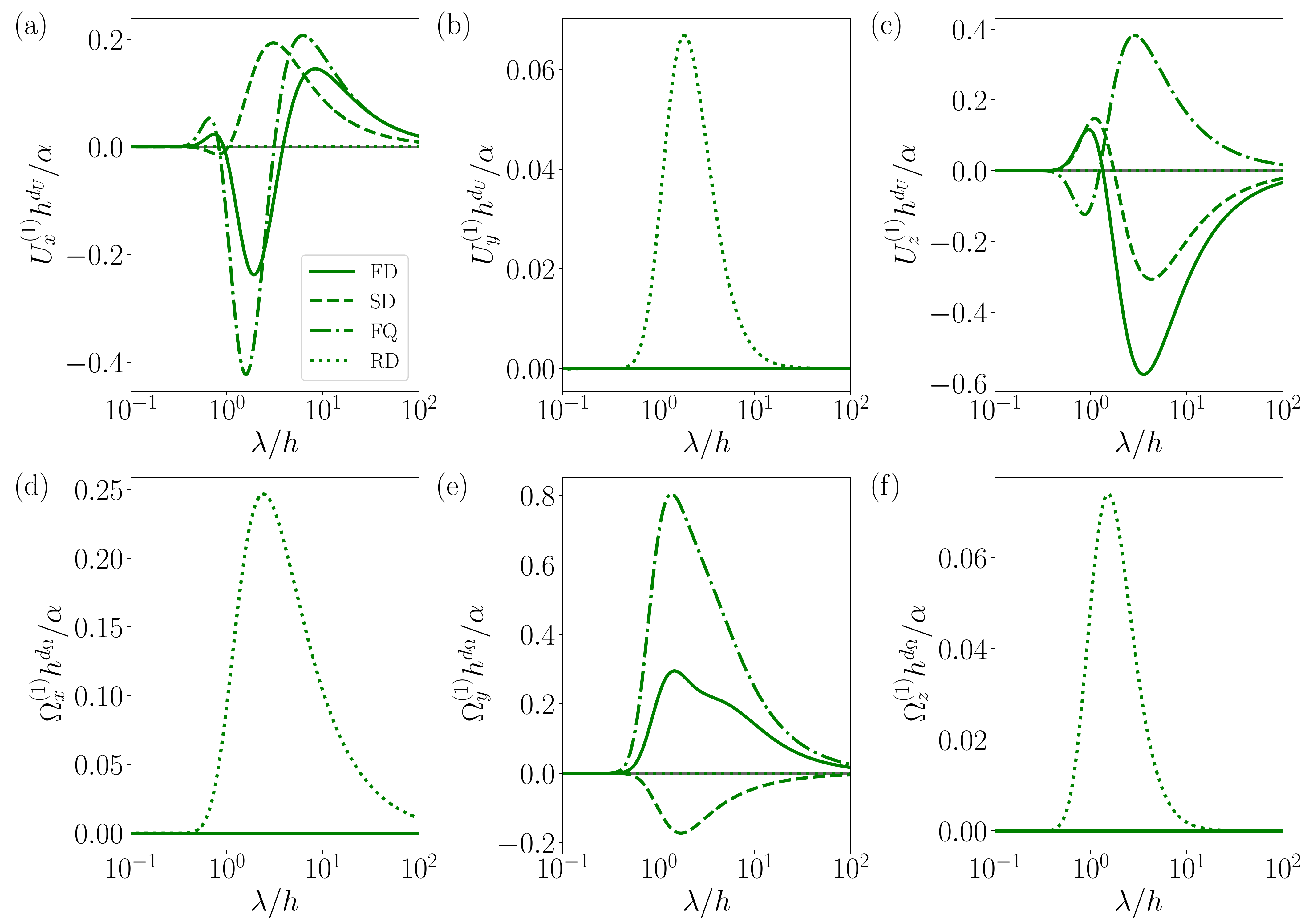}
\caption{Roughness-induced velocities of a force dipole (FD), source dipole (SD), force quadrupole (FQ), and rotlet dipole (RD) near a wavy surface of shape $H(x,y)=\cos(2\pi x/\lambda)$. The components of the velocities, $\vec{U}^{(1)}$ (a-c) and $\boldsymbol{\Omega}^{(1)}$ (d-f), are shown with respect to the wavelength $\lambda/h$ for a microswimmer located at a distance $h/a=2$, at position $x_S/\lambda = 0.25$, and with pitch angle $\vartheta_0=-\pi/8$ and $\varphi_0=0$. Here, $d_U=2$ ($d_U=3$) and $d_\Omega=3$ ($d_\Omega=4$) for the force dipole (higher-order singularities) and $\alpha =\alpha_{FD}, \alpha_{SD},\alpha_{FQ},\alpha_{RD}$, respectively.
\label{fig:velocities_x025}}
\end{figure*}

\subsection{Periodic surface shape}
We consider the roughness-induced velocities of Stokes singularities near a periodic, wavy surface, $H(x,y)=\cos(2\pi x/\lambda)$, where $\lambda$ denotes the characteristic wavelength. The calculations of the velocities can be mostly performed analytically up to the radial integral, which we then evaluate numerically. As example, we present the case of a force dipole in the Appendix~\ref{appendix:rough}.

Since roughness-induced velocities depend locally on the particle-surface configuration, we selected two cases to illustrate the role of the characteristic wavelength of the surface on the translational and rotational velocities. In particular, varying the wavelength $\lambda$ effectively changes the slope of the surface located below the microswimmer as well as the width of the valley that modifies the hydrodynamic flows. Figures~\ref{fig:velocities_x0} and \ref{fig:velocities_x025} show the results for a microswimmer directed perpendicular to the surface grooves ($\varphi_0=0$) at an angle $\vartheta_0=-\pi/8$ towards the periodic surface for varying $\lambda/h$ and for different swimmer positions $\vec{r}_0= [x_S, y_S, h]^T$. In particular, we fix the swimmer-surface distance $h/a=2$, set $y_S=0$, and consider $x_S/\lambda = 0$ and  $x_S/\lambda = 0.25$.

For wavelengths much smaller than the particle distance from the wall, $0.1\lesssim\lambda/h\lesssim1$, roughness-induced velocities approach zero. In this case, the surface area closest to the agent contains several surface bumps
that smear out roughness-induced flows and therefore the microswimmer experiences the average wall contribution only. This behavior could change for the case where the surface tips are longer than the wavelength $\lambda\lesssim\epsilon h$. Then the tips may dominate the overall surface contribution. This, however, is not captured by our domain perturbation method and further analysis about its validity for small $\lambda$ is required.

For large wavelengths, $\lambda/h\gg1$, and for a microswimmer located on top of a hill or a valley (i.e., $x_S/\lambda  =0, 0.5$) the surface essentially appears closer to or further away from the swimmer, $h\to h(1\mp \epsilon)$, whereas the surface shape does not impact the velocities since its slope becomes negligible. An expansion in $\epsilon$ indicates that the velocities for a force dipole generated by the presence of a planar wall are proportional to
\begin{align}
U^*\propto \frac{\alpha_{FD}}{h^2(1\mp\epsilon)^2} = \frac{\alpha_{FD}}{h^2}(1\pm2\epsilon)+\mathcal{O}(\epsilon^2)
\end{align}
and similarly for higher-order singularities, $U^*\propto\alpha(1\pm3\epsilon)/h^3+\mathcal{O}(\epsilon^2)$.
Thus, the roughness-induced velocities at $x_S/\lambda = 0, 0.5$ approach $\vec{U}^\epsilon = \pm d_U \vec{U}^{*}$, where $d_U=2$ for a force dipole and $d_U=3$ for higher-order singularities (SD, FQ, RD). The angular velocities tend towards $\boldsymbol{\Omega}^\epsilon = \pm d_{\Omega} \boldsymbol{\Omega}^{*}$, with $d_{\Omega}=3$ for a force dipole and $d_\Omega=4$ for the higher-order singularities. We observe that the roughness-induced velocities approach those induced by a shifted planar wall at $\lambda/h\gtrsim10$ [Fig.~\ref{fig:velocities_x0}].

Moreover, the roughness-induced velocities of a particle located at $x_S/\lambda  =0.25$ (and $x_S/\lambda  =0.75$ (not shown)) vanish for large wavelengths, $\lambda/h\gg1$, as the underlying surface shape, $z=\epsilon h H(x,y)$, approaches the reference surface, $S_0$ (see Fig.~\ref{fig:velocities_x025}).
The translational and angular velocities induced by a wavy surface, $\vec{U}^{(1)}$ and $\boldsymbol{\Omega}^{(1)}$, display non-trivial behavior at wavelengths $1\lesssim \lambda/h\lesssim 10$, where the hydrodynamic couplings of the self-propelled particle and the boundary depend on details of the surface topography.

In particular, the contribution due to roughness can be hydrodynamically attractive or repulsive depending on the wavelength of the surface. A pusher-type microswimmer (at $x_S/\lambda = 0, 0.25$) experiences a repulsive contribution from the surface, $U_z^{(1)}>0$, at wavelengths $\lambda/h\sim1$, which indicates that the extensile flow-field becomes reflected from the underlying cavity and thereby pushes the swimmer away from the surface. At larger wavelengths the pusher becomes even more attracted to the surface, $U_z^{(1)}<U_z^\epsilon$, since the cavity provides enough extra space for the incoming fluid flow (see Figs.~\ref{fig:velocities_x0}(c) and~\ref{fig:velocities_x025}(c)).

The inverse effect occurs for a puller with $\alpha_{FD}<0$, where a surface cavity of length $\lambda/h\sim1$ contributes attractive rather than repulsive interactions. Higher-order singularities (SD and FQ) also induce an inverse behavior of $U_z^{(1)}$ at wavelengths $\lambda/h\sim1$ compared to the contribution of a planar wall.

Particle velocities along the $x-$direction of pushers (FD) above a surface of wavelength $\lambda/h\sim1$ are enhanced on top of a hill, $U_x^{(1)}>0$, whereas at $x_S/\lambda=0.25$ they are decreased, $U_x^{(1)}<0$ (see Figs.~\ref{fig:velocities_x0}(a) and~\ref{fig:velocities_x025}(a)). However, for larger wavelengths $\lambda/h\sim10$ the velocity contribution due to the roughness becomes positive $U_x^{(1)}>0$ at $x_S/\lambda=0.25$, which indicates an enhancement due to a large underlying surface cavity.

Due to the symmetry of the surface along the  $y-$direction, a force dipole, source dipole, and force quadrupole do not contribute to the roughness-induced velocities, $U_y^{(1)}$. However, a rotlet dipole, which induces clockwise circular motion for $\alpha_{RD}>0$, displays non-vanishing velocities at $1\lesssim \lambda/h\lesssim 10$ (see Figs.~\ref{fig:velocities_x0}(b) and~\ref{fig:velocities_x025}(b)). In particular, these are negative at $x_S/\lambda=0$ and positive at $x_S/\lambda=0.25$, where the interaction with the surface induces motion opposite to the direction of rotation. Similarly, at these wavelengths the roughness-induced angular velocity becomes positive, $\Omega_z^{(1)}>0$, and thereby reduces the clockwise rotation (see Figs.~\ref{fig:velocities_x0}(f) and~\ref{fig:velocities_x025}(f)). The rotlet dipole also generates rotation around~$\vec{e}_x$ (see Figs.~\ref{fig:velocities_x0}(d) and~\ref{fig:velocities_x025}(d)).

Near a planar wall, a pusher with pitch angle $\vartheta_0=-\pi/8$ tends to align parallel to the surface and thus $\Omega^*_y<0$ (see Eq.~\eqref{eq:omega_fd}). The angular velocities $\Omega_y^{(1)}$ induced by a wavy wall indicate a similar effect that is determined by the slope of the underlying surface (see Fig.~\ref{fig:velocities_x025}(e)). In particular, at $x_S/\lambda=0.25$ the roughness-induced velocities contribute with $\Omega_y^{(1)}>0$ and thereby indicate alignment parallel with the (steeper) surface slope.

Whether a puller ($\alpha_{FD}<0$) rotates towards or away from the surface crucially depends on the pitch angle. In particular, near a planar wall it rotates away for $\vartheta_0>0$ and towards the surface for $\vartheta_0<0$ (see Eq.~\eqref{eq:omega_fd}). However, at $x_S/\lambda=0.25$ we find that for a puller with pitch angle $\vartheta_0=-\pi/8$ the first-order contribution due to the wavy surface is negative $\Omega_y^{(1)}<0$ and hence contributes to a rotation away from the surface. This effect due to the wavy surface decreases the overall rotation towards the surface (as $\Omega_y^{*}>0$).

\begin{figure}[htp]
    \centering
    \includegraphics[width = \linewidth]{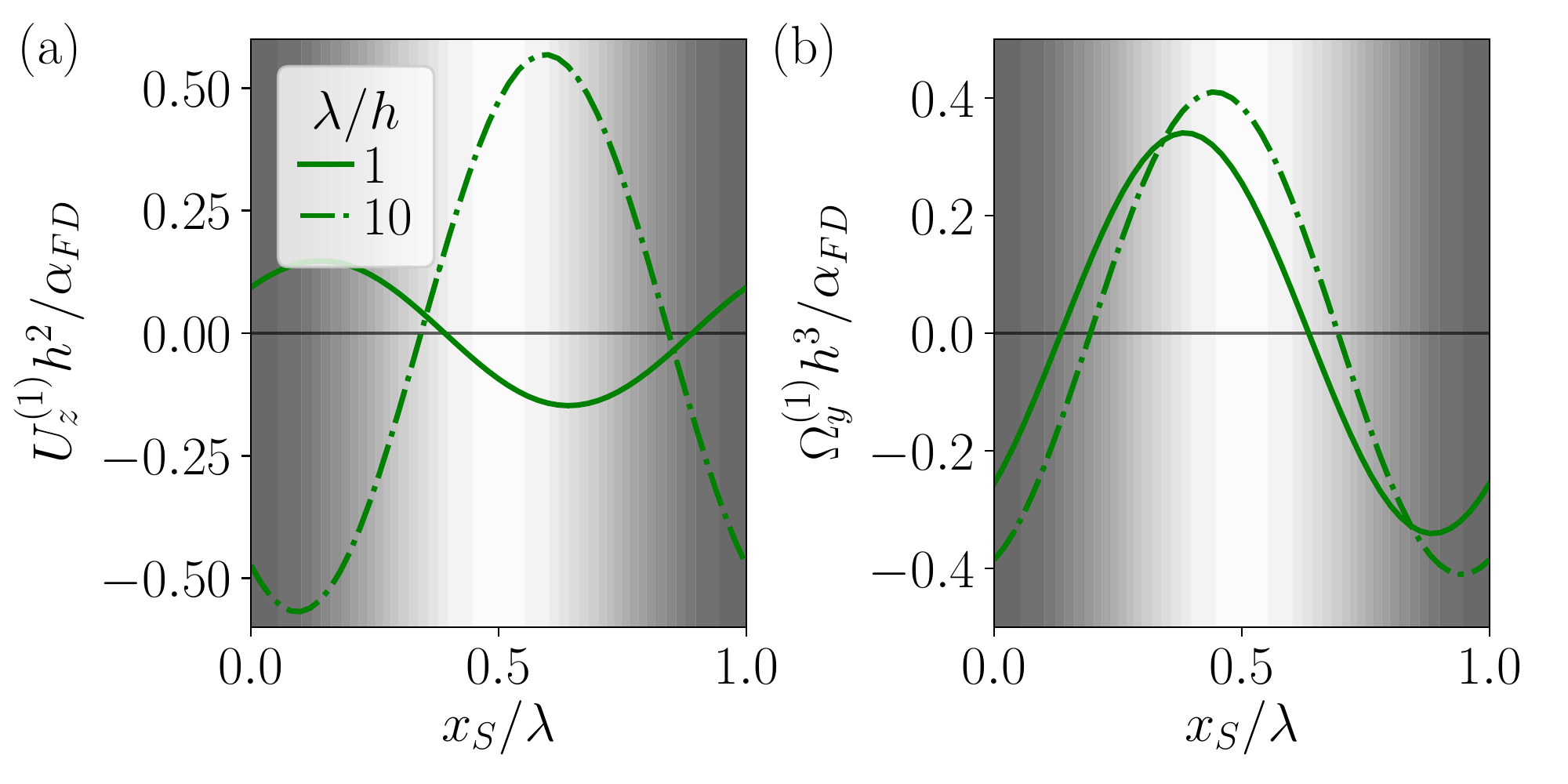}
    \caption{Components of roughness-induced velocities of a microswimmer, (a) $U^{(1)}_z$ and (b) $\Omega^{(1)}_y$, as a function of the particle position~$x_S$ for different wavelengths~$\lambda$. The swimmer is modeled as force dipole (FD) located at $h/a=2$ with $\vartheta_0 = -\pi/8$ and $\varphi_0=0$. The gray shaded areas indicate the underlying surface height: dark areas correspond to hills and light areas to cavities.}
    \label{fig:6}
\end{figure}

Finally, we provide results for fixed wavelengths $\lambda$ as a function of the position $x_S$. We consider the components $U_z^{(1)}$ and $\Omega_y^{(1)}$ for a swimmer modeled as a force dipole with pitch angle $\vartheta_0=-\pi/8$ and distance $h/a=2$, see Fig.~\ref{fig:6}. We observe that for a wavelength comparable to the distance from the surface, $\lambda/h=1$, the roughness causes a pusher to be attracted to cavities and repelled from hills [Fig.~\ref{fig:6}(a)]. This, however, changes for larger wavelengths ($\lambda/h= 10$), where close to hills the hydrodynamic attraction becomes enhanced, while it becomes decreased near cavities. The rotational velocities indicate to promote rotation of a pusher so that its swimming direction aligns parallel to the surface slope: it rotates away from the wall at $x_S/\lambda\sim 0.5$, while it rotates towards the wall for larger and smaller $x_S$ [Fig.~\ref{fig:6}(b)].  These results remain largely unaffected by changing $\lambda$. The opposite holds for pullers.

\subsection{\emph{E. coli} near a wavy surface\label{sec:ecoli}} 
\begin{figure}[htp]
     \centering
     \includegraphics[width = 0.9\linewidth]{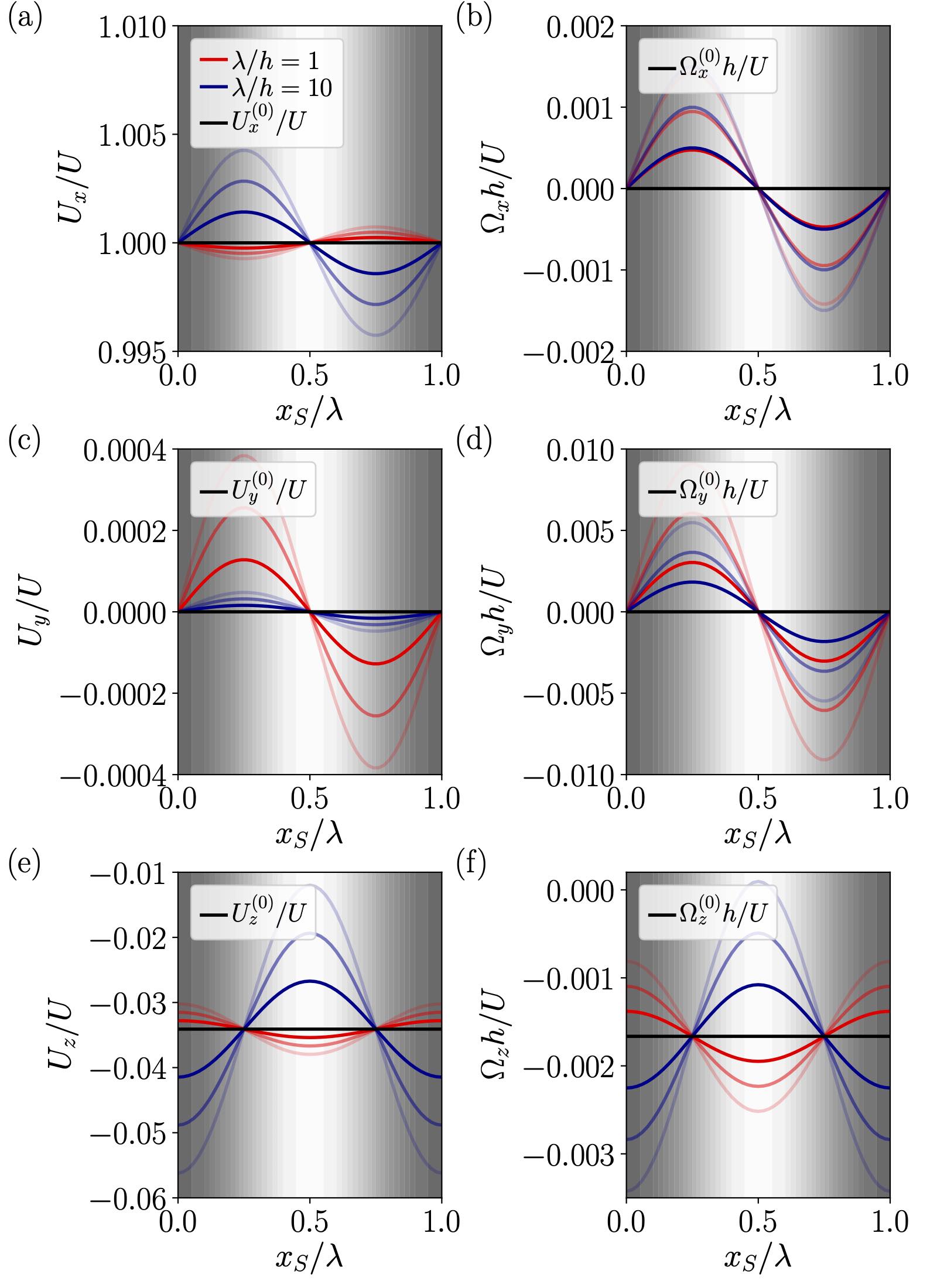}
     \caption{Velocities, $\vec{U}= \vec{U}^{(0)}+\epsilon\vec{U}^{(1)}$ and $\boldsymbol{\Omega}= \boldsymbol{\Omega}^{(0)}+\epsilon\boldsymbol{\Omega}^{(1)}$, of an \emph{E. coli} bacterium located at a distance $h=4\, \mu$m with orientation $\vartheta_0=0$ near a periodic surface. Red and blue indicate the wavelengths $\lambda/h = 1$ and $\lambda/h = 10$, respectively. Different opacities indicate different roughnesses ($\epsilon = 0.1,0.2,0.3$): from $\epsilon = 0.1$ (dark) to $\epsilon = 0.3$ (light). Further, $U=22\,\mu$m$\cdot$s$^{-1}$ is its swim speed in an unconfined environment and black lines denote the velocities, $\vec{U}^{(0)}$ and $\boldsymbol{\Omega}^{(0)}$, near a planar wall. The gray shaded areas indicate the underlying surface height: dark areas correspond to hills and light areas to cavities.}
     \label{fig:7}
 \end{figure}

To investigate the effect of surface roughness on the velocities of particular microswimmers, the contributions due to the individual Stokes singularities can be added up using experimentally-measured singularity strengths. For example, the singularity strength of a force dipole field produced by an \emph{E. coli} bacterium, which swims at speed  $U=22\, \mu$m$\cdot$s$^{-1}$, has been measured experimentally~\cite{Drescher:2010}. In this study, the dipole force and the dipole length yielded, respectively, $F \simeq 0.42$~pN and $\ell \simeq 1.9\,\mu$m.
A theoretical study~\cite{Hu:2015:SM} has corroborated these findings using simulations of flagellated bacteria. In addition, it provided the torque of the rotlet dipole, $M\simeq80 k_BT$ with Boltzmann constant $k_B$ and temperature $T$, which is generated by the rotation of the flagellum and the counter-rotation of the cell body. Assuming the same length $\ell$ for the rotlet dipole and using the viscosity of water, provides an estimate for the force dipole and rotlet dipole strengths:  $\alpha_{FD}=F\ell/(8\pi\mu)\simeq32\, \mu$m$^3\cdot$s$^{-1}$ and $\alpha_{RD}= M\ell/(8\pi\mu)\simeq25\, \mu$m$^4\cdot$s$^{-1}$.
We use these as inputs to study the translational and angular velocities of an \emph{E. coli} bacterium with orientation $\vartheta_0=0$ near a wavy surface with different wavelengths~$\lambda$ and roughness~$\epsilon$, see Fig.~\ref{fig:7}.

As our theory is linear in the surface roughness, increasing $\epsilon$ increases the contribution due to the corrugated surface shape. However, as discussed earlier, different wavelengths can change qualitatively the contributions of the surface shape. This becomes manifest most prominently in $U_z$ and $\Omega_z$ [ Fig.~\ref{fig:7}(e),(f)]. In particular, for $\lambda/h =10$ the bacterium is attracted by hills more than valleys and the clockwise swimming motion becomes enhanced at hills compared to valleys. A surface roughness of $\epsilon = 0.3$ in fact indicates a counter-clockwise rotation on top of a valley opposed to the clockwise rotation near a planar wall. For a bacterium moving near the wavy surface this could lead to overall clockwise swimming motion with oscillations. This behavior changes for a surface with a smaller wavelength $\lambda/h = 1$. The roughness-induced contributions to $\Omega_z$ and $U_y$ originate from the rotlet dipole flow.

We further observe that the microswimmer with $\vartheta_0 = 0$, which remains aligned parallel to the planar wall ($\Omega_y=0$), tends to align with the slope of the corrugated surface. More specifically, it rotates towards the surface ($\Omega_y>0$) for $0<x_s/\lambda<0.5$ and away from it for $0.5<x_s/\lambda<1$. Also, the velocities parallel, $U_x$, and transverse to the wall, $U_y$, become enhanced on top of inclined downhills while they decrease near upwards slopes for $\lambda/h=10$. For $\lambda/h=1$ this behavior becomes more pronounced for the transverse velocity, $U_y$, but reverses and becomes much smaller for $U_x$. The surface shape induces rotation around the long axis of the swimming bacterium, $\vec{e}_x$, which depends on the surface slope.

Finally, we note that the contributions of the surface shape resulting from hydrodynamic coupling remain of the order of $10^{-4}-10^{-2}$ of the swim speed in an unconfined environment. However, our analysis is limited to small surface roughness and valid in a far-field description only and, thus, the effects are expected to become more pronounced in the near-field limit. The results also depend on the orientation of the swimmer. While we have limited the discussion here to a swimmer aligned parallel to the surface, the results for an \emph{E. coli} bacterium with orientation $\vartheta_0=-\pi/8$ are shown in the appendix~\ref{appendix:ecoli} [Fig.~\ref{fig:8}].

\section{Summary and conclusion}
\begin{figure}[htp]
\centering
\includegraphics[width=\linewidth]{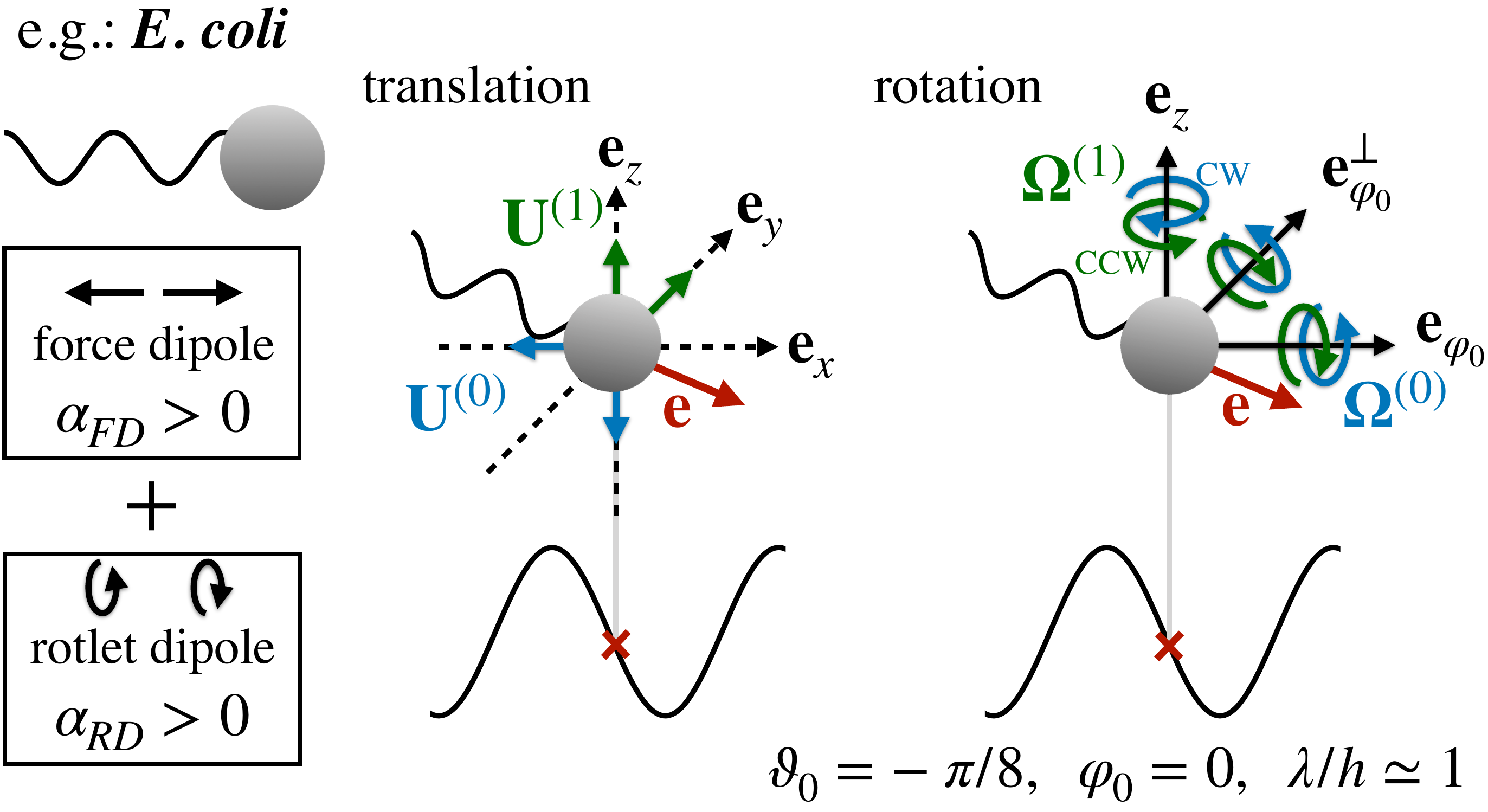}
\caption{Visualization of our qualitative conclusions (with results from Fig.~\ref{fig:velocities_x025}) for an \emph{E. coli} bacterium modeled as a superposition of a force dipole and rotlet dipole. Arrows merely indicate the sign of the velocities near a planar wall, $\vec{U}^{(0)}$ and $\boldsymbol{\Omega}^{(0)}$, and roughness-induced contributions, $\vec{U}^{(1)}$ and $\boldsymbol{\Omega}^{(1)}$. Here, CCW and CW denote counter-clockwise and clockwise rotation, respectively.
\label{fig:sketch_results}}
\end{figure}

We have presented analytical expressions for the roughness-induced velocities of microswimmers near textured surfaces characterized by arbitrary shapes up to first order in the surface amplitude. We have applied our theoretical predictions to study the effect of a wavy surface on the velocities of microswimmers modeled in terms of a multipole expansion, where we have accounted for flow fields of a force dipole, source dipole, force quadrupole, and rotlet dipole.
Our results show that surface cavities, which are comparable in size with the particle distance from the wall, can produce repulsive contributions to the velocities of a pusher as the extensile flow fields are reflected at the edge of the cavity. Furthermore, the clockwise circular swimming motion of, e.g., \emph{E. coli}, bacteria near a wall~\cite{Lauga:2006} is affected by the wavy surface shape, which can contribute to a counter-clockwise sense of rotation or enhance the clockwise rotation depending on the surface wavelength and its location with respect to surface hills or valleys (see Fig.~\ref{fig:sketch_results} for a visualization of our conclusions).

Our findings, valid in the framework of a far-field analysis, $h/a\gg 1$, suggest that surface cavities can possibly decrease accumulation of pusher-type microswimmers or enhance surface accumulation of pullers. Since the swimming direction is also affected by the underlying surface, our theory indicates that hydrodynamic interactions can indeed contribute to a randomization of deterministic circular swimming motion of bacteria nearby planar walls, as has been observed experimentally~\cite{Makarchuk:2019}.
Yet, details of these possible behaviors remain to be elucidated by accounting for the near-wall lubrication flows.

The attachment of different types of swimming and non-swimming cells near corrugated channels in shear flow has been investigated experimentally. These experiments have shown preferred leeward attachment of \emph{E. coli} at curved elements of the surface, while passive cells prefer attachment at the windward side closer to the surface peaks~\cite{Secchi:2020}. 
In the future it will be interesting to study how external flows affect our findings and thereby provide insights into the microscopic dynamics of the experimental observations.
The surface effect should become most pronounced in the close vicinity to the surface, where details of the flows close to the microswimmer body become important. As the flow fields generated by a squirming sphere near a planar wall have been elaborated analytically~\cite{Shaik:2017}, our theory can readily be applied to elucidate squirming motion near textured walls and to study the effect of surface heterogeneities on the dynamical behavior. In particular, a superposition of periodic modes can be used to represent (random) rough surface shapes~\cite{Kamrin:2010, Kurzthaler:2020}, which could shed light on microswimmer motion near realistic biological surfaces.

\section*{Conflicts of interest}
There are no conflicts to declare.


\begin{appendix}
\section{Appendix: Multipole expansion\label{appendix}}
\subsection{Translational and rotational velocities induced by a planar wall\label{appendix:vel}}
The velocities induced by the presence of a smooth, planar wall~\cite{Spagnolie:2012} are presented here for the source dipole,
\begin{subequations}
\begin{align}
\vec{U}^*_{SD}&= -\frac{\alpha_{SD}}{4h^3}\cos\vartheta_0\vec{e}_{\varphi_0}-\frac{\alpha_{SD}}{h^3}\sin\vartheta_0\vec{e}_z,\\
\boldsymbol{\Omega}^*_{SD}&= -\frac{3\alpha_{SD}}{8h^4}\cos\vartheta_0\vec{e}_{\varphi_0}^\perp,
\end{align}
\end{subequations}
the force quadrupole,
\begin{subequations}
\begin{align}
\begin{split}
\vec{U}^*_{FQ}&= \frac{\alpha_{FQ}}{32h^3}\cos\vartheta_0(-13+27\cos(2\vartheta_0))\vec{e}_{\varphi_0}+\\
& \ \ \ \ \ \frac{\alpha_{FQ}}{16h^3}(\sin\vartheta_0+9\sin(3\vartheta_0))\vec{e}_z,
\end{split}\\
\boldsymbol{\Omega}^*_{FQ}&=\frac{3\alpha_{FQ}}{32h^4}(\cos\vartheta_0+3\cos(3\vartheta_0))\vec{e}_{\varphi_0}^\perp,
\end{align}
\end{subequations}
and the rotlet dipole, $\vec{U}^*_{RD}=\vec{0}$ and
\begin{align}
\begin{split}
\boldsymbol{\Omega}^*_{RD}=&\frac{9\alpha_{RD}}{32h^4}\sin(2\vartheta_0) \vec{e}_{\varphi_0}\\
&-\frac{3\alpha_{RD}}{64h^4}(-1+3\cos(2\vartheta_0))\vec{e}_z.
\end{split}
\end{align}
\subsection{Calculation of roughness-induced velocities\label{appendix:rough}}
Here, we explicitly show the calculation of the roughness-induced velocities of a swimmer modeled as force dipole with velocity field $\vec{u}^{(0)}=\vec{u}_{FD}+\vec{u}_{FD}^*$ near a planar wall. The dimensionless components for the velocity gradient at the no-slip wall (Z=0) are
\begin{subequations}
\begin{align}
\begin{split}
    &\left[\frac{\partial U_R^{(0)}}{\partial Z}\right]_{Z=0} =\\
    & \quad \frac{3}{(1 + R^2)^{7/2}}\Bigl(2 R (4 R^2-1) \cos(2(\varphi - \varphi_0)) \cos^2\vartheta_0 +\\
    & \quad R (2 R^2-3)(3 \cos(2 \vartheta_0)-1) +\\
   & \quad 2 (1 - 8 R^2 + R^4) \cos(\varphi - \varphi_0) \sin(2 \vartheta_0)\Bigr),
   \end{split}\label{uz_1}\\
\begin{split}&\left[\frac{\partial U_\varphi^{(0)}}{\partial Z}\right]_{Z=0} =\\
    & \frac{12 \cos\vartheta_0\sin(\varphi - \varphi_0) (R \cos(\varphi - \varphi_0) \cos\vartheta_0 - \sin\vartheta_0)}{(1 + R^2)^{5/2}}.
\end{split}\label{uz_2}
\end{align}
\end{subequations}
To evaluate the roughness-induced velocities we require the stresses of the auxiliary problem. The stresses of a stokeslet, which is located at the singularity position $\vec{r}_0$ and directed along $\vec{e}_x$, $\vec{e}_y$, or $\vec{e}_z$, at the no-slip wall ($Z=0$) are:
\begin{subequations}
\begin{align}
    \left[\hat{\Sigma}^{U,x}_{ZR}\right]_{Z=0} &= \frac{3R^2 \cos\varphi}{2\pi(1 + R^2)^{5/2}}, \\
    \left[\hat{\Sigma}^{U,y}_{ZR}\right]_{Z=0} &= \frac{3R^2 \sin\varphi}{2\pi(1 + R^2)^{5/2}}, \\
    \left[\hat{\Sigma}^{U,z}_{ZR}\right]_{Z=0} &= -\frac{3R}{2\pi(1 + R^2)^{5/2}}.
\end{align}
\end{subequations}
Similarly, the stresses induced by rotlets evaluate to
\begin{subequations}
\begin{align}
    \left[\hat{\Sigma}^{\Omega,x}_{ZR}\right]_{Z=0}\!\! &= -\frac{3 (R^2-1) \sin\varphi}{4 \pi (1 + R^2)^{5/2}},\\
     \left[\hat{\Sigma}^{\Omega,x}_{Z\varphi}\right]_{Z=0}\!\! &= \frac{3\cos\varphi}{4 \pi (1 + R^2)^{5/2}},\\
    \left[\hat{\Sigma}^{\Omega,y}_{ZR}\right]_{Z=0}\!\! &= \frac{3 ( R^2-1) \cos\varphi}{4 \pi (1 + R^2)^{5/2}},\\
     \left[\hat{\Sigma}^{\Omega,y}_{Z\varphi}\right]_{Z=0}\!\! &= \frac{3\sin\varphi}{4 \pi (1 + R^2)^{5/2}},\\
    \left[\hat{\Sigma}^{\Omega,z}_{ZR}\right]_{Z=0}\!\! &= 0,\\
    \left[\hat{\Sigma}^{\Omega,z}_{Z\varphi}\right]_{Z=0}\!\! &= \frac{3 R}{4 \pi (1 + R^2)^{5/2}}. \label{sigma}
\end{align}
\end{subequations}
We note that the presence of the wall leads to the generation of hydrodynamic torques and, thus, rotation of the forced point particle, which we have neglected here as they do not contribute to leading order in $h$. Using the expressions [Eqs.~\eqref{uz_1}-\eqref{sigma}] as input for Eqs.~\eqref{eq:U}-\eqref{eq:O} provides the roughness-induced velocities of a microswimmer modeled as a force dipole. As an example, we calculate the roughness-induced translational velocity along the $z-$direction:
\begin{widetext}
\begin{align}
\begin{split}
U^{(1)}_z = -\frac{\alpha_{FD}}{h^2}\int_0^\infty \frac{9 R^2}{\left(R^2+1\right)^6} \times 
\Biggl[&2 R \left(4 R^2-1\right) J_2\left(\frac{2 \pi h R}{\lambda}\right) \cos (2 \varphi_0) \cos ^2\vartheta_0 \cos \left(\frac{2 \pi  x_S}{\lambda}\right)+ \\
&2 \left(R^4-8 R^2+1\right)J_1\left(\frac{2 \pi h R}{\lambda}\right) \cos\varphi_0 \sin (2 \vartheta_0) \sin \left(\frac{2 \pi  x_S}{\lambda}\right)-\\
&R \left(2 R^2-3\right)J_0\left(\frac{2 \pi h  R}{\lambda}\right) (3 \cos (2 \vartheta_0)-1) \cos \left(\frac{2 \pi  x_S}{\lambda}\right)\Biggr] \ \diff R,
\end{split}
\end{align}
\end{widetext}
where $J_n(\cdot)$ denotes the Bessel function of order $n$ and the final radial integral is performed numerically. Also, we present the $y-$component of the roughness-induced rotational velocity:
\begin{widetext}
\begin{align}
\begin{split}
\Omega^{(1)}_y = \frac{\alpha_{FD}}{h^3}\int_0^\infty\frac{9}{2 \pi  \left(R^2+1\right)^6}  \Biggl[&2 R J_2\left(\frac{2 h \pi  R}{\lambda}\right) \times
\Biggl(\pi  \left(R^6-9 R^4+9 R^2-1\right) \cos\varphi_0 \sin (2 \vartheta_0) \cos \left(\frac{2 \pi  x_S}{\lambda}\right)+\\
&2 \frac{\lambda}{h} R^2 \left(3-2 R^2\right) \cos (2 \varphi_0) \cos ^2\vartheta_0 \sin \left(\frac{2 \pi  x_S}{\lambda}\right)\Biggr)+\\
&J_1\left(\frac{2\pi h  R}{\lambda}\right) \Biggl(\pi  R^2 \left(R^2-1\right) \sin \left(\frac{2 \pi  x_S}{\lambda}\right)\times \notag\\
&\left(2 \left(4 R^2-1\right) \cos (2 \varphi_0) \cos ^2\vartheta_0+\left(2 R^2-3\right) (3 \cos (2 \vartheta_0)-1)\right)-\\
&\frac{\lambda}{h} \left(R^6-9 R^4+8 R^2-2\right) \cos\varphi_0 \sin (2 \vartheta_0) \cos \left(\frac{2 \pi  x_S}{\lambda}\right)\Biggr)\Biggr] \ \diff R.
\end{split}
\end{align}
\end{widetext}

To calculate the roughness-induced velocities of higher-order singularities (SD, FQ, and RD) the velocity gradient [Eqs.~\eqref{uz_1}-\eqref{uz_2}] has to be amplified (see Ref.~\cite{Spagnolie:2012}). Additional details on the calculations can be provided upon request.

\subsection{\emph{E. coli} near a wavy surface with orientation $\vartheta_0=-\pi/8$\label{appendix:ecoli}}
\begin{figure}[htp]
     \centering
     \includegraphics[width = 0.9\linewidth]{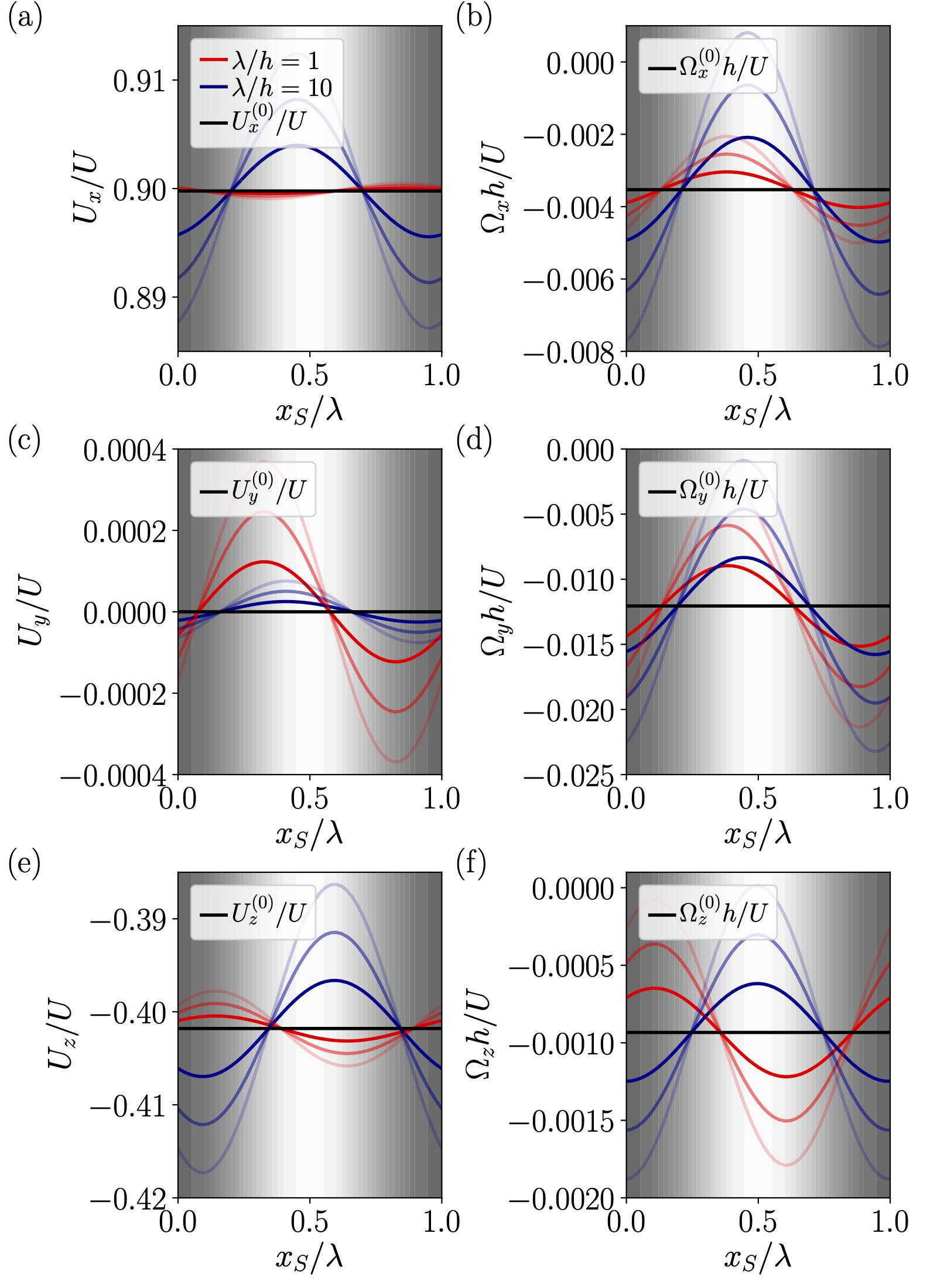}
     \caption{Velocities, $\vec{U}= \vec{U}^{(0)}+\epsilon\vec{U}^{(1)}$ and $\boldsymbol{\Omega}= \boldsymbol{\Omega}^{(0)}+\epsilon\boldsymbol{\Omega}^{(1)}$, of an \emph{E. coli} bacterium located at a distance $h=4\, \mu$m with orientation $\vartheta_0=-\pi/8$ near a periodic surface. Red and blue indicate the wavelengths $\lambda/h = 1$ and $\lambda/h = 10$, respectively. Different opacities indicate different roughnesses ($\epsilon = 0.1,0.2,0.3$): from $\epsilon = 0.1$ (dark) to $\epsilon = 0.3$ (light). Further, $U=22\,\mu$m$\cdot$s$^{-1}$ is its swim speed in an unconfined environment and black lines denote the velocities, $\vec{U}^{(0)}$ and $\boldsymbol{\Omega}^{(0)}$, near a planar wall. The gray shaded areas indicate the underlying surface height: dark areas correspond to hills and light areas to cavities.\label{fig:8} }
 \end{figure}

Figure~\ref{fig:8} shows the results of an \emph{E. coli} bacterium with swimming direction $\vartheta_0 = -\pi/8$. The motility parameters are discussed in Sec.~\ref{sec:ecoli}.

 \end{appendix}

 \section*{Acknowledgements}
C.K. acknowledges support from the Austrian Science Fund (FWF) via the Erwin Schr{\"o}dinger fellowship (Grant No. J4321-N27). The work was supported by NSF MCB-1853602 (H.A.S.).
\bibliography{literature} 

\begin{thebibliography}{60}%
\makeatletter
\providecommand \@ifxundefined [1]{%
 \@ifx{#1\undefined}
}%
\providecommand \@ifnum [1]{%
 \ifnum #1\expandafter \@firstoftwo
 \else \expandafter \@secondoftwo
 \fi
}%
\providecommand \@ifx [1]{%
 \ifx #1\expandafter \@firstoftwo
 \else \expandafter \@secondoftwo
 \fi
}%
\providecommand \natexlab [1]{#1}%
\providecommand \enquote  [1]{``#1''}%
\providecommand \bibnamefont  [1]{#1}%
\providecommand \bibfnamefont [1]{#1}%
\providecommand \citenamefont [1]{#1}%
\providecommand \href@noop [0]{\@secondoftwo}%
\providecommand \href [0]{\begingroup \@sanitize@url \@href}%
\providecommand \@href[1]{\@@startlink{#1}\@@href}%
\providecommand \@@href[1]{\endgroup#1\@@endlink}%
\providecommand \@sanitize@url [0]{\catcode `\\12\catcode `\$12\catcode
  `\&12\catcode `\#12\catcode `\^12\catcode `\_12\catcode `\%12\relax}%
\providecommand \@@startlink[1]{}%
\providecommand \@@endlink[0]{}%
\providecommand \url  [0]{\begingroup\@sanitize@url \@url }%
\providecommand \@url [1]{\endgroup\@href {#1}{\urlprefix }}%
\providecommand \urlprefix  [0]{URL }%
\providecommand \Eprint [0]{\href }%
\providecommand \doibase [0]{https://doi.org/}%
\providecommand \selectlanguage [0]{\@gobble}%
\providecommand \bibinfo  [0]{\@secondoftwo}%
\providecommand \bibfield  [0]{\@secondoftwo}%
\providecommand \translation [1]{[#1]}%
\providecommand \BibitemOpen [0]{}%
\providecommand \bibitemStop [0]{}%
\providecommand \bibitemNoStop [0]{.\EOS\space}%
\providecommand \EOS [0]{\spacefactor3000\relax}%
\providecommand \BibitemShut  [1]{\csname bibitem#1\endcsname}%
\let\auto@bib@innerbib\@empty
\bibitem [{\citenamefont {Höfling}\ and\ \citenamefont
  {Franosch}(2013)}]{Hoefling:2013}%
  \BibitemOpen
  \bibfield  {author} {\bibinfo {author} {\bibfnamefont {F.}~\bibnamefont
  {Höfling}}\ and\ \bibinfo {author} {\bibfnamefont {T.}~\bibnamefont
  {Franosch}},\ }\bibfield  {title} {\bibinfo {title} {Anomalous transport in
  the crowded world of biological cells},\ }\href
  {https://doi.org/10.1088/0034-4885/76/4/046602} {\bibfield  {journal}
  {\bibinfo  {journal} {Rep. Prog. Phys.}\ }\textbf {\bibinfo {volume} {76}},\
  \bibinfo {pages} {046602} (\bibinfo {year} {2013})}\BibitemShut {NoStop}%
\bibitem [{\citenamefont {Persat}\ \emph {et~al.}(2015)\citenamefont {Persat},
  \citenamefont {Nadell}, \citenamefont {Kim}, \citenamefont {Ingremeau},
  \citenamefont {Siryaporn}, \citenamefont {Drescher}, \citenamefont
  {Wingreen}, \citenamefont {Bassler}, \citenamefont {Gitai},\ and\
  \citenamefont {Stone}}]{Persat:2015}%
  \BibitemOpen
  \bibfield  {author} {\bibinfo {author} {\bibfnamefont {A.}~\bibnamefont
  {Persat}}, \bibinfo {author} {\bibfnamefont {C.~D.}\ \bibnamefont {Nadell}},
  \bibinfo {author} {\bibfnamefont {M.~K.}\ \bibnamefont {Kim}}, \bibinfo
  {author} {\bibfnamefont {F.}~\bibnamefont {Ingremeau}}, \bibinfo {author}
  {\bibfnamefont {A.}~\bibnamefont {Siryaporn}}, \bibinfo {author}
  {\bibfnamefont {K.}~\bibnamefont {Drescher}}, \bibinfo {author}
  {\bibfnamefont {N.~S.}\ \bibnamefont {Wingreen}}, \bibinfo {author}
  {\bibfnamefont {B.~L.}\ \bibnamefont {Bassler}}, \bibinfo {author}
  {\bibfnamefont {Z.}~\bibnamefont {Gitai}},\ and\ \bibinfo {author}
  {\bibfnamefont {H.~A.}\ \bibnamefont {Stone}},\ }\bibfield  {title} {\bibinfo
  {title} {The mechanical world of bacteria},\ }\href
  {https://doi.org/10.1016/j.cell.2015.05.005} {\bibfield  {journal} {\bibinfo
  {journal} {Cell}\ }\textbf {\bibinfo {volume} {161}},\ \bibinfo {pages} {988}
  (\bibinfo {year} {2015})}\BibitemShut {NoStop}%
\bibitem [{\citenamefont {Bechinger}\ \emph {et~al.}(2016)\citenamefont
  {Bechinger}, \citenamefont {Di~Leonardo}, \citenamefont {L\"owen},
  \citenamefont {Reichhardt}, \citenamefont {Volpe},\ and\ \citenamefont
  {Volpe}}]{Bechinger:2016}%
  \BibitemOpen
  \bibfield  {author} {\bibinfo {author} {\bibfnamefont {C.}~\bibnamefont
  {Bechinger}}, \bibinfo {author} {\bibfnamefont {R.}~\bibnamefont
  {Di~Leonardo}}, \bibinfo {author} {\bibfnamefont {H.}~\bibnamefont
  {L\"owen}}, \bibinfo {author} {\bibfnamefont {C.}~\bibnamefont {Reichhardt}},
  \bibinfo {author} {\bibfnamefont {G.}~\bibnamefont {Volpe}},\ and\ \bibinfo
  {author} {\bibfnamefont {G.}~\bibnamefont {Volpe}},\ }\bibfield  {title}
  {\bibinfo {title} {Active particles in complex and crowded environments},\
  }\href {https://doi.org/10.1103/RevModPhys.88.045006} {\bibfield  {journal}
  {\bibinfo  {journal} {Rev. Mod. Phys.}\ }\textbf {\bibinfo {volume} {88}},\
  \bibinfo {pages} {045006} (\bibinfo {year} {2016})}\BibitemShut {NoStop}%
\bibitem [{\citenamefont {Berne}\ \emph {et~al.}(2018)\citenamefont {Berne},
  \citenamefont {Ellison}, \citenamefont {Ducret},\ and\ \citenamefont
  {Brun}}]{Berne:2018}%
  \BibitemOpen
  \bibfield  {author} {\bibinfo {author} {\bibfnamefont {C.}~\bibnamefont
  {Berne}}, \bibinfo {author} {\bibfnamefont {C.~K.}\ \bibnamefont {Ellison}},
  \bibinfo {author} {\bibfnamefont {A.}~\bibnamefont {Ducret}},\ and\ \bibinfo
  {author} {\bibfnamefont {Y.~V.}\ \bibnamefont {Brun}},\ }\bibfield  {title}
  {\bibinfo {title} {Bacterial adhesion at the single-cell level},\ }\href@noop
  {} {\bibfield  {journal} {\bibinfo  {journal} {Nat. Rev. Microbiol.}\
  }\textbf {\bibinfo {volume} {16}},\ \bibinfo {pages} {616} (\bibinfo {year}
  {2018})}\BibitemShut {NoStop}%
\bibitem [{\citenamefont {Bhattacharjee}\ and\ \citenamefont
  {Datta}(2019)}]{Bhattacharjee:2019}%
  \BibitemOpen
  \bibfield  {author} {\bibinfo {author} {\bibfnamefont {T.}~\bibnamefont
  {Bhattacharjee}}\ and\ \bibinfo {author} {\bibfnamefont {S.~S.}\ \bibnamefont
  {Datta}},\ }\bibfield  {title} {\bibinfo {title} {Bacterial hopping and
  trapping in porous media},\ }\href
  {https://doi.org/10.1038/s41467-019-10115-1} {\bibfield  {journal} {\bibinfo
  {journal} {Nat. {C}ommun.}\ }\textbf {\bibinfo {volume} {10}},\ \bibinfo
  {pages} {2075} (\bibinfo {year} {2019})}\BibitemShut {NoStop}%
\bibitem [{\citenamefont {Daddi-Moussa-Ider}\ \emph {et~al.}(2019)\citenamefont
  {Daddi-Moussa-Ider}, \citenamefont {Kurzthaler}, \citenamefont {Hoell},
  \citenamefont {Z\"ottl}, \citenamefont {Mirzakhanloo}, \citenamefont {Alam},
  \citenamefont {Menzel}, \citenamefont {L\"owen},\ and\ \citenamefont
  {Gekle}}]{Daddi:2019}%
  \BibitemOpen
  \bibfield  {author} {\bibinfo {author} {\bibfnamefont {A.}~\bibnamefont
  {Daddi-Moussa-Ider}}, \bibinfo {author} {\bibfnamefont {C.}~\bibnamefont
  {Kurzthaler}}, \bibinfo {author} {\bibfnamefont {C.}~\bibnamefont {Hoell}},
  \bibinfo {author} {\bibfnamefont {A.}~\bibnamefont {Z\"ottl}}, \bibinfo
  {author} {\bibfnamefont {M.}~\bibnamefont {Mirzakhanloo}}, \bibinfo {author}
  {\bibfnamefont {M.-R.}\ \bibnamefont {Alam}}, \bibinfo {author}
  {\bibfnamefont {A.~M.}\ \bibnamefont {Menzel}}, \bibinfo {author}
  {\bibfnamefont {H.}~\bibnamefont {L\"owen}},\ and\ \bibinfo {author}
  {\bibfnamefont {S.}~\bibnamefont {Gekle}},\ }\bibfield  {title} {\bibinfo
  {title} {Frequency-dependent higher-order stokes singularities near a planar
  elastic boundary: Implications for the hydrodynamics of an active
  microswimmer near an elastic interface},\ }\href
  {https://doi.org/10.1103/PhysRevE.100.032610} {\bibfield  {journal} {\bibinfo
   {journal} {Phys. Rev. E}\ }\textbf {\bibinfo {volume} {100}},\ \bibinfo
  {pages} {032610} (\bibinfo {year} {2019})}\BibitemShut {NoStop}%
\bibitem [{\citenamefont {Uspal}\ \emph {et~al.}(2019)\citenamefont {Uspal},
  \citenamefont {Popescu}, \citenamefont {Dietrich},\ and\ \citenamefont
  {Tasinkevych}}]{Uspal:2019}%
  \BibitemOpen
  \bibfield  {author} {\bibinfo {author} {\bibfnamefont {W.~E.}\ \bibnamefont
  {Uspal}}, \bibinfo {author} {\bibfnamefont {M.~N.}\ \bibnamefont {Popescu}},
  \bibinfo {author} {\bibfnamefont {S.}~\bibnamefont {Dietrich}},\ and\
  \bibinfo {author} {\bibfnamefont {M.}~\bibnamefont {Tasinkevych}},\
  }\bibfield  {title} {\bibinfo {title} {Active janus colloids at chemically
  structured surfaces},\ }\href {https://doi.org/10.1063/1.5091760} {\bibfield
  {journal} {\bibinfo  {journal} {J. Chem. Phys.}\ }\textbf {\bibinfo {volume}
  {150}},\ \bibinfo {pages} {204904} (\bibinfo {year} {2019})}\BibitemShut
  {NoStop}%
\bibitem [{\citenamefont {Poortinga}\ \emph {et~al.}(2002)\citenamefont
  {Poortinga}, \citenamefont {Bos}, \citenamefont {Norde},\ and\ \citenamefont
  {Busscher}}]{Poortinga:2002}%
  \BibitemOpen
  \bibfield  {author} {\bibinfo {author} {\bibfnamefont {A.~T.}\ \bibnamefont
  {Poortinga}}, \bibinfo {author} {\bibfnamefont {R.}~\bibnamefont {Bos}},
  \bibinfo {author} {\bibfnamefont {W.}~\bibnamefont {Norde}},\ and\ \bibinfo
  {author} {\bibfnamefont {H.~J.}\ \bibnamefont {Busscher}},\ }\bibfield
  {title} {\bibinfo {title} {Electric double layer interactions in bacterial
  adhesion to surfaces},\ }\href@noop {} {\bibfield  {journal} {\bibinfo
  {journal} {Surf. Sci. Rep.}\ }\textbf {\bibinfo {volume} {47}},\ \bibinfo
  {pages} {1} (\bibinfo {year} {2002})}\BibitemShut {NoStop}%
\bibitem [{\citenamefont {Humphries}\ \emph {et~al.}(2017)\citenamefont
  {Humphries}, \citenamefont {Xiong}, \citenamefont {Liu}, \citenamefont
  {Prindle}, \citenamefont {Yuan}, \citenamefont {Arjes}, \citenamefont
  {Tsimring},\ and\ \citenamefont {S{\"u}el}}]{Humphries:2017}%
  \BibitemOpen
  \bibfield  {author} {\bibinfo {author} {\bibfnamefont {J.}~\bibnamefont
  {Humphries}}, \bibinfo {author} {\bibfnamefont {L.}~\bibnamefont {Xiong}},
  \bibinfo {author} {\bibfnamefont {J.}~\bibnamefont {Liu}}, \bibinfo {author}
  {\bibfnamefont {A.}~\bibnamefont {Prindle}}, \bibinfo {author} {\bibfnamefont
  {F.}~\bibnamefont {Yuan}}, \bibinfo {author} {\bibfnamefont {H.~A.}\
  \bibnamefont {Arjes}}, \bibinfo {author} {\bibfnamefont {L.}~\bibnamefont
  {Tsimring}},\ and\ \bibinfo {author} {\bibfnamefont {G.~M.}\ \bibnamefont
  {S{\"u}el}},\ }\bibfield  {title} {\bibinfo {title} {Species-independent
  attraction to biofilms through electrical signaling},\ }\href@noop {}
  {\bibfield  {journal} {\bibinfo  {journal} {Cell}\ }\textbf {\bibinfo
  {volume} {168}},\ \bibinfo {pages} {200} (\bibinfo {year}
  {2017})}\BibitemShut {NoStop}%
\bibitem [{\citenamefont {Kroy}\ \emph {et~al.}(2016)\citenamefont {Kroy},
  \citenamefont {Chakraborty},\ and\ \citenamefont {Cichos}}]{Kroy:2016}%
  \BibitemOpen
  \bibfield  {author} {\bibinfo {author} {\bibfnamefont {K.}~\bibnamefont
  {Kroy}}, \bibinfo {author} {\bibfnamefont {D.}~\bibnamefont {Chakraborty}},\
  and\ \bibinfo {author} {\bibfnamefont {F.}~\bibnamefont {Cichos}},\
  }\bibfield  {title} {\bibinfo {title} {Hot microswimmers},\ }\href@noop {}
  {\bibfield  {journal} {\bibinfo  {journal} {The European Physical Journal
  Special Topics}\ }\textbf {\bibinfo {volume} {225}},\ \bibinfo {pages} {2207}
  (\bibinfo {year} {2016})}\BibitemShut {NoStop}%
\bibitem [{\citenamefont {Lushi}\ \emph {et~al.}(2017)\citenamefont {Lushi},
  \citenamefont {Kantsler},\ and\ \citenamefont {Goldstein}}]{Lushi:2017}%
  \BibitemOpen
  \bibfield  {author} {\bibinfo {author} {\bibfnamefont {E.}~\bibnamefont
  {Lushi}}, \bibinfo {author} {\bibfnamefont {V.}~\bibnamefont {Kantsler}},\
  and\ \bibinfo {author} {\bibfnamefont {R.~E.}\ \bibnamefont {Goldstein}},\
  }\bibfield  {title} {\bibinfo {title} {Scattering of biflagellate
  microswimmers from surfaces},\ }\href
  {https://doi.org/10.1103/PhysRevE.96.023102} {\bibfield  {journal} {\bibinfo
  {journal} {Phys. Rev. E}\ }\textbf {\bibinfo {volume} {96}},\ \bibinfo
  {pages} {023102} (\bibinfo {year} {2017})}\BibitemShut {NoStop}%
\bibitem [{\citenamefont {Leal}(2007)}]{Leal:2007}%
  \BibitemOpen
  \bibfield  {author} {\bibinfo {author} {\bibfnamefont {L.~G.}\ \bibnamefont
  {Leal}},\ }\href@noop {} {\emph {\bibinfo {title} {Advanced {T}ransport
  {P}henomena: {F}luid {M}echanics and {C}onvective {T}ransport
  {P}rocesses}}},\ Vol.~\bibinfo {volume} {7}\ (\bibinfo  {publisher}
  {Cambridge University Press},\ \bibinfo {year} {2007})\BibitemShut {NoStop}%
\bibitem [{\citenamefont {Berke}\ \emph {et~al.}(2008)\citenamefont {Berke},
  \citenamefont {Turner}, \citenamefont {Berg},\ and\ \citenamefont
  {Lauga}}]{Berke:2008}%
  \BibitemOpen
  \bibfield  {author} {\bibinfo {author} {\bibfnamefont {A.~P.}\ \bibnamefont
  {Berke}}, \bibinfo {author} {\bibfnamefont {L.}~\bibnamefont {Turner}},
  \bibinfo {author} {\bibfnamefont {H.~C.}\ \bibnamefont {Berg}},\ and\
  \bibinfo {author} {\bibfnamefont {E.}~\bibnamefont {Lauga}},\ }\bibfield
  {title} {\bibinfo {title} {Hydrodynamic attraction of swimming microorganisms
  by surfaces},\ }\href {https://doi.org/10.1103/PhysRevLett.101.038102}
  {\bibfield  {journal} {\bibinfo  {journal} {Phys. Rev. Lett.}\ }\textbf
  {\bibinfo {volume} {101}},\ \bibinfo {pages} {038102} (\bibinfo {year}
  {2008})}\BibitemShut {NoStop}%
\bibitem [{\citenamefont {Berg}\ and\ \citenamefont {Brown}(1972)}]{Berg:1972}%
  \BibitemOpen
  \bibfield  {author} {\bibinfo {author} {\bibfnamefont {H.~C.}\ \bibnamefont
  {Berg}}\ and\ \bibinfo {author} {\bibfnamefont {D.~A.}\ \bibnamefont
  {Brown}},\ }\bibfield  {title} {\bibinfo {title} {Chemotaxis in escherichia
  coli analysed by three-dimensional tracking},\ }\href
  {https://doi.org/10.1038/239500a0} {\bibfield  {journal} {\bibinfo  {journal}
  {Nature}\ }\textbf {\bibinfo {volume} {239}},\ \bibinfo {pages} {500}
  (\bibinfo {year} {1972})}\BibitemShut {NoStop}%
\bibitem [{\citenamefont {Berg}(2008)}]{Berg:2008}%
  \BibitemOpen
  \bibfield  {author} {\bibinfo {author} {\bibfnamefont {H.~C.}\ \bibnamefont
  {Berg}},\ }\href {https://doi.org/10.1007/b97370} {\emph {\bibinfo {title}
  {E. coli in Motion}}},\ Biological and Medical Physics, Biomedical
  Engineering\ (\bibinfo  {publisher} {Springer Science and Business Media, New
  York},\ \bibinfo {year} {2008})\BibitemShut {NoStop}%
\bibitem [{\citenamefont {Woolley}(2003)}]{Woolley:2003}%
  \BibitemOpen
  \bibfield  {author} {\bibinfo {author} {\bibfnamefont {D.}~\bibnamefont
  {Woolley}},\ }\bibfield  {title} {\bibinfo {title} {Motility of spermatozoa
  at surfaces},\ }\href {https://doi.org/10.1530/rep.0.1260259} {\bibfield
  {journal} {\bibinfo  {journal} {Reprod.}\ }\textbf {\bibinfo {volume}
  {126}},\ \bibinfo {pages} {259} (\bibinfo {year} {2003})}\BibitemShut
  {NoStop}%
\bibitem [{\citenamefont {Riedel}\ \emph {et~al.}(2005)\citenamefont {Riedel},
  \citenamefont {Kruse},\ and\ \citenamefont {Howard}}]{Riedel:2005}%
  \BibitemOpen
  \bibfield  {author} {\bibinfo {author} {\bibfnamefont {I.~H.}\ \bibnamefont
  {Riedel}}, \bibinfo {author} {\bibfnamefont {K.}~\bibnamefont {Kruse}},\ and\
  \bibinfo {author} {\bibfnamefont {J.}~\bibnamefont {Howard}},\ }\bibfield
  {title} {\bibinfo {title} {A self-organized vortex array of hydrodynamically
  entrained sperm cells},\ }\href {https://doi.org/10.1126/science.1110329}
  {\bibfield  {journal} {\bibinfo  {journal} {Science}\ }\textbf {\bibinfo
  {volume} {309}},\ \bibinfo {pages} {300} (\bibinfo {year}
  {2005})}\BibitemShut {NoStop}%
\bibitem [{\citenamefont {Friedrich}\ and\ \citenamefont
  {J{\"u}licher}(2008)}]{Friedrich:2008}%
  \BibitemOpen
  \bibfield  {author} {\bibinfo {author} {\bibfnamefont {B.}~\bibnamefont
  {Friedrich}}\ and\ \bibinfo {author} {\bibfnamefont {F.}~\bibnamefont
  {J{\"u}licher}},\ }\bibfield  {title} {\bibinfo {title} {The stochastic dance
  of circling sperm cells: sperm chemotaxis in the plane},\ }\href
  {https://doi.org/10.1088/1367-2630/10/12/123025} {\bibfield  {journal}
  {\bibinfo  {journal} {New J. Phys.}\ }\textbf {\bibinfo {volume} {10}},\
  \bibinfo {pages} {123025} (\bibinfo {year} {2008})}\BibitemShut {NoStop}%
\bibitem [{\citenamefont {Machemer}(1972)}]{Machemer:1972}%
  \BibitemOpen
  \bibfield  {author} {\bibinfo {author} {\bibfnamefont {H.}~\bibnamefont
  {Machemer}},\ }\bibfield  {title} {\bibinfo {title} {Ciliary activity and the
  origin of metachrony in paramecium: effects of increased viscosity},\ }\href
  {http://jeb.biologists.org/content/57/1/239} {\bibfield  {journal} {\bibinfo
  {journal} {J. Exp. Biol.}\ }\textbf {\bibinfo {volume} {57}},\ \bibinfo
  {pages} {239} (\bibinfo {year} {1972})}\BibitemShut {NoStop}%
\bibitem [{\citenamefont {Merchant~\textit{et al.}}(2007)}]{Merchant:2007}%
  \BibitemOpen
  \bibfield  {author} {\bibinfo {author} {\bibfnamefont {S.~S.}\ \bibnamefont
  {Merchant~\textit{et al.}}},\ }\bibfield  {title} {\bibinfo {title} {The
  chlamydomonas genome reveals the evolution of key animal and plant
  functions},\ }\href {https://doi.org/10.1126/science.1143609} {\bibfield
  {journal} {\bibinfo  {journal} {Science}\ }\textbf {\bibinfo {volume}
  {318}},\ \bibinfo {pages} {245} (\bibinfo {year} {2007})}\BibitemShut
  {NoStop}%
\bibitem [{\citenamefont {Howse}\ \emph {et~al.}(2007)\citenamefont {Howse},
  \citenamefont {Jones}, \citenamefont {Ryan}, \citenamefont {Gough},
  \citenamefont {Vafabakhsh},\ and\ \citenamefont {Golestanian}}]{Howse:2007}%
  \BibitemOpen
  \bibfield  {author} {\bibinfo {author} {\bibfnamefont {J.~R.}\ \bibnamefont
  {Howse}}, \bibinfo {author} {\bibfnamefont {R.~A.~L.}\ \bibnamefont {Jones}},
  \bibinfo {author} {\bibfnamefont {A.~J.}\ \bibnamefont {Ryan}}, \bibinfo
  {author} {\bibfnamefont {T.}~\bibnamefont {Gough}}, \bibinfo {author}
  {\bibfnamefont {R.}~\bibnamefont {Vafabakhsh}},\ and\ \bibinfo {author}
  {\bibfnamefont {R.}~\bibnamefont {Golestanian}},\ }\bibfield  {title}
  {\bibinfo {title} {Self-motile colloidal particles: From directed propulsion
  to random walk},\ }\href {https://doi.org/10.1103/PhysRevLett.99.048102}
  {\bibfield  {journal} {\bibinfo  {journal} {Phys. Rev. Lett.}\ }\textbf
  {\bibinfo {volume} {99}},\ \bibinfo {pages} {048102} (\bibinfo {year}
  {2007})}\BibitemShut {NoStop}%
\bibitem [{\citenamefont {Buttinoni}\ \emph {et~al.}(2012)\citenamefont
  {Buttinoni}, \citenamefont {Volpe}, \citenamefont {Kümmel}, \citenamefont
  {Volpe},\ and\ \citenamefont {Bechinger}}]{Buttinoni:2012}%
  \BibitemOpen
  \bibfield  {author} {\bibinfo {author} {\bibfnamefont {I.}~\bibnamefont
  {Buttinoni}}, \bibinfo {author} {\bibfnamefont {G.}~\bibnamefont {Volpe}},
  \bibinfo {author} {\bibfnamefont {F.}~\bibnamefont {Kümmel}}, \bibinfo
  {author} {\bibfnamefont {G.}~\bibnamefont {Volpe}},\ and\ \bibinfo {author}
  {\bibfnamefont {C.}~\bibnamefont {Bechinger}},\ }\bibfield  {title} {\bibinfo
  {title} {Active brownian motion tunable by light},\ }\href
  {https://doi.org/10.1088/0953-8984/24/28/284129} {\bibfield  {journal}
  {\bibinfo  {journal} {J. Condens. Matter Phys.}\ }\textbf {\bibinfo {volume}
  {24}},\ \bibinfo {pages} {284129} (\bibinfo {year} {2012})}\BibitemShut
  {NoStop}%
\bibitem [{\citenamefont {Kurzthaler}\ \emph {et~al.}(2018)\citenamefont
  {Kurzthaler}, \citenamefont {Devailly}, \citenamefont {Arlt}, \citenamefont
  {Franosch}, \citenamefont {Poon}, \citenamefont {Martinez},\ and\
  \citenamefont {Brown}}]{Kurzthaler:2018}%
  \BibitemOpen
  \bibfield  {author} {\bibinfo {author} {\bibfnamefont {C.}~\bibnamefont
  {Kurzthaler}}, \bibinfo {author} {\bibfnamefont {C.}~\bibnamefont
  {Devailly}}, \bibinfo {author} {\bibfnamefont {J.}~\bibnamefont {Arlt}},
  \bibinfo {author} {\bibfnamefont {T.}~\bibnamefont {Franosch}}, \bibinfo
  {author} {\bibfnamefont {W.~C.~K.}\ \bibnamefont {Poon}}, \bibinfo {author}
  {\bibfnamefont {V.~A.}\ \bibnamefont {Martinez}},\ and\ \bibinfo {author}
  {\bibfnamefont {A.~T.}\ \bibnamefont {Brown}},\ }\bibfield  {title} {\bibinfo
  {title} {Probing the spatiotemporal dynamics of catalytic janus particles
  with single-particle tracking and differential dynamic microscopy},\ }\href
  {https://doi.org/10.1103/PhysRevLett.121.078001} {\bibfield  {journal}
  {\bibinfo  {journal} {Phys. Rev. Lett.}\ }\textbf {\bibinfo {volume} {121}},\
  \bibinfo {pages} {078001} (\bibinfo {year} {2018})}\BibitemShut {NoStop}%
\bibitem [{\citenamefont {Dreyfus}\ \emph {et~al.}(2005)\citenamefont
  {Dreyfus}, \citenamefont {Baudry}, \citenamefont {Roper}, \citenamefont
  {Fermigier}, \citenamefont {Stone},\ and\ \citenamefont
  {Bibette}}]{Dreyfus:2005}%
  \BibitemOpen
  \bibfield  {author} {\bibinfo {author} {\bibfnamefont {R.}~\bibnamefont
  {Dreyfus}}, \bibinfo {author} {\bibfnamefont {J.}~\bibnamefont {Baudry}},
  \bibinfo {author} {\bibfnamefont {M.~L.}\ \bibnamefont {Roper}}, \bibinfo
  {author} {\bibfnamefont {M.}~\bibnamefont {Fermigier}}, \bibinfo {author}
  {\bibfnamefont {H.~A.}\ \bibnamefont {Stone}},\ and\ \bibinfo {author}
  {\bibfnamefont {J.}~\bibnamefont {Bibette}},\ }\bibfield  {title} {\bibinfo
  {title} {Microscopic artificial swimmers},\ }\href
  {https://doi.org/10.1038/nature04090} {\bibfield  {journal} {\bibinfo
  {journal} {Nature}\ }\textbf {\bibinfo {volume} {437}},\ \bibinfo {pages}
  {862} (\bibinfo {year} {2005})}\BibitemShut {NoStop}%
\bibitem [{\citenamefont {Ghosh}\ and\ \citenamefont
  {Fischer}(2009)}]{Ghosh:2009}%
  \BibitemOpen
  \bibfield  {author} {\bibinfo {author} {\bibfnamefont {A.}~\bibnamefont
  {Ghosh}}\ and\ \bibinfo {author} {\bibfnamefont {P.}~\bibnamefont
  {Fischer}},\ }\bibfield  {title} {\bibinfo {title} {Controlled propulsion of
  artificial magnetic nanostructured propellers},\ }\href
  {https://doi.org/10.1021/nl900186w} {\bibfield  {journal} {\bibinfo
  {journal} {Nano Lett.}\ }\textbf {\bibinfo {volume} {9}},\ \bibinfo {pages}
  {2243} (\bibinfo {year} {2009})}\BibitemShut {NoStop}%
\bibitem [{\citenamefont {O'Neill}\ and\ \citenamefont
  {Stewartson}(1967)}]{Oneill:1967}%
  \BibitemOpen
  \bibfield  {author} {\bibinfo {author} {\bibfnamefont {M.}~\bibnamefont
  {O'Neill}}\ and\ \bibinfo {author} {\bibfnamefont {K.}~\bibnamefont
  {Stewartson}},\ }\bibfield  {title} {\bibinfo {title} {On the slow motion of
  a sphere parallel to a nearby plane wall},\ }\href
  {https://doi.org/10.1017/S0022112067002551} {\bibfield  {journal} {\bibinfo
  {journal} {J. Fluid Mech.}\ }\textbf {\bibinfo {volume} {27}},\ \bibinfo
  {pages} {705} (\bibinfo {year} {1967})}\BibitemShut {NoStop}%
\bibitem [{\citenamefont {Goldman}\ \emph {et~al.}(1967)\citenamefont
  {Goldman}, \citenamefont {Cox},\ and\ \citenamefont
  {Brenner}}]{Goldman:1967}%
  \BibitemOpen
  \bibfield  {author} {\bibinfo {author} {\bibfnamefont {A.~J.}\ \bibnamefont
  {Goldman}}, \bibinfo {author} {\bibfnamefont {R.~G.}\ \bibnamefont {Cox}},\
  and\ \bibinfo {author} {\bibfnamefont {H.}~\bibnamefont {Brenner}},\
  }\bibfield  {title} {\bibinfo {title} {Slow viscous motion of a sphere
  parallel to a plane wall—i motion through a quiescent fluid},\ }\href
  {https://doi.org/10.1016/0009-2509(67)80047-2} {\bibfield  {journal}
  {\bibinfo  {journal} {Chem. Eng. Sci.}\ }\textbf {\bibinfo {volume} {22}},\
  \bibinfo {pages} {637} (\bibinfo {year} {1967})}\BibitemShut {NoStop}%
\bibitem [{\citenamefont {Zöttl}\ and\ \citenamefont
  {Stark}(2016)}]{Zoettl:2016}%
  \BibitemOpen
  \bibfield  {author} {\bibinfo {author} {\bibfnamefont {A.}~\bibnamefont
  {Zöttl}}\ and\ \bibinfo {author} {\bibfnamefont {H.}~\bibnamefont {Stark}},\
  }\bibfield  {title} {\bibinfo {title} {Emergent behavior in active
  colloids},\ }\href {https://doi.org/10.1088/0953-8984/28/25/253001}
  {\bibfield  {journal} {\bibinfo  {journal} {J. Condens. Matter Phys.}\
  }\textbf {\bibinfo {volume} {28}},\ \bibinfo {pages} {253001} (\bibinfo
  {year} {2016})}\BibitemShut {NoStop}%
\bibitem [{\citenamefont {Drescher}\ \emph {et~al.}(2011)\citenamefont
  {Drescher}, \citenamefont {Dunkel}, \citenamefont {Cisneros}, \citenamefont
  {Ganguly},\ and\ \citenamefont {Goldstein}}]{Drescher:2011}%
  \BibitemOpen
  \bibfield  {author} {\bibinfo {author} {\bibfnamefont {K.}~\bibnamefont
  {Drescher}}, \bibinfo {author} {\bibfnamefont {J.}~\bibnamefont {Dunkel}},
  \bibinfo {author} {\bibfnamefont {L.~H.}\ \bibnamefont {Cisneros}}, \bibinfo
  {author} {\bibfnamefont {S.}~\bibnamefont {Ganguly}},\ and\ \bibinfo {author}
  {\bibfnamefont {R.~E.}\ \bibnamefont {Goldstein}},\ }\bibfield  {title}
  {\bibinfo {title} {Fluid dynamics and noise in bacterial
  cell{\textendash}cell and cell{\textendash}surface scattering},\ }\href
  {https://doi.org/10.1073/pnas.1019079108} {\bibfield  {journal} {\bibinfo
  {journal} {Proc. Natl. Acad. Sci. U. S. A.}\ }\textbf {\bibinfo {volume}
  {108}},\ \bibinfo {pages} {10940} (\bibinfo {year} {2011})}\BibitemShut
  {NoStop}%
\bibitem [{\citenamefont {Drescher}\ \emph {et~al.}(2010)\citenamefont
  {Drescher}, \citenamefont {Goldstein}, \citenamefont {Michel}, \citenamefont
  {Polin},\ and\ \citenamefont {Tuval}}]{Drescher:2010}%
  \BibitemOpen
  \bibfield  {author} {\bibinfo {author} {\bibfnamefont {K.}~\bibnamefont
  {Drescher}}, \bibinfo {author} {\bibfnamefont {R.~E.}\ \bibnamefont
  {Goldstein}}, \bibinfo {author} {\bibfnamefont {N.}~\bibnamefont {Michel}},
  \bibinfo {author} {\bibfnamefont {M.}~\bibnamefont {Polin}},\ and\ \bibinfo
  {author} {\bibfnamefont {I.}~\bibnamefont {Tuval}},\ }\bibfield  {title}
  {\bibinfo {title} {Direct measurement of the flow field around swimming
  microorganisms},\ }\href {https://doi.org/10.1103/PhysRevLett.105.168101}
  {\bibfield  {journal} {\bibinfo  {journal} {Phys. Rev. Lett.}\ }\textbf
  {\bibinfo {volume} {105}},\ \bibinfo {pages} {168101} (\bibinfo {year}
  {2010})}\BibitemShut {NoStop}%
\bibitem [{\citenamefont {Uspal}\ \emph {et~al.}(2015)\citenamefont {Uspal},
  \citenamefont {Popescu}, \citenamefont {Dietrich},\ and\ \citenamefont
  {Tasinkevych}}]{Uspal:2015}%
  \BibitemOpen
  \bibfield  {author} {\bibinfo {author} {\bibfnamefont {W.~E.}\ \bibnamefont
  {Uspal}}, \bibinfo {author} {\bibfnamefont {M.~N.}\ \bibnamefont {Popescu}},
  \bibinfo {author} {\bibfnamefont {S.}~\bibnamefont {Dietrich}},\ and\
  \bibinfo {author} {\bibfnamefont {M.}~\bibnamefont {Tasinkevych}},\
  }\bibfield  {title} {\bibinfo {title} {Self-propulsion of a catalytically
  active particle near a planar wall: from reflection to sliding and
  hovering},\ }\href {https://doi.org/10.1039/C4SM02317J} {\bibfield  {journal}
  {\bibinfo  {journal} {Soft Matter}\ }\textbf {\bibinfo {volume} {11}},\
  \bibinfo {pages} {434} (\bibinfo {year} {2015})}\BibitemShut {NoStop}%
\bibitem [{\citenamefont {Shen}\ \emph {et~al.}(2018)\citenamefont {Shen},
  \citenamefont {W{\"u}rger},\ and\ \citenamefont {Lintuvuori}}]{Shen:2018}%
  \BibitemOpen
  \bibfield  {author} {\bibinfo {author} {\bibfnamefont {Z.}~\bibnamefont
  {Shen}}, \bibinfo {author} {\bibfnamefont {A.}~\bibnamefont {W{\"u}rger}},\
  and\ \bibinfo {author} {\bibfnamefont {J.~S.}\ \bibnamefont {Lintuvuori}},\
  }\bibfield  {title} {\bibinfo {title} {Hydrodynamic interaction of a
  self-propelling particle with a wall},\ }\href
  {https://doi.org/10.1140/epje/i2018-11649-0} {\bibfield  {journal} {\bibinfo
  {journal} {Eur. Phys. J. E Soft Matter}\ }\textbf {\bibinfo {volume} {41}},\
  \bibinfo {pages} {39} (\bibinfo {year} {2018})}\BibitemShut {NoStop}%
\bibitem [{\citenamefont {Fauci}\ and\ \citenamefont
  {McDonald}(1995)}]{Fauci:1995}%
  \BibitemOpen
  \bibfield  {author} {\bibinfo {author} {\bibfnamefont {L.~J.}\ \bibnamefont
  {Fauci}}\ and\ \bibinfo {author} {\bibfnamefont {A.}~\bibnamefont
  {McDonald}},\ }\bibfield  {title} {\bibinfo {title} {Sperm motility in the
  presence of boundaries},\ }\href {https://doi.org/10.1007/BF02461846}
  {\bibfield  {journal} {\bibinfo  {journal} {Bull. Math. Biol.}\ }\textbf
  {\bibinfo {volume} {57}},\ \bibinfo {pages} {679} (\bibinfo {year}
  {1995})}\BibitemShut {NoStop}%
\bibitem [{\citenamefont {Smith}\ \emph {et~al.}(2009)\citenamefont {Smith},
  \citenamefont {Gaffney}, \citenamefont {Blake},\ and\ \citenamefont
  {Kirkman-Brown}}]{Smith:2009}%
  \BibitemOpen
  \bibfield  {author} {\bibinfo {author} {\bibfnamefont {D.~J.}\ \bibnamefont
  {Smith}}, \bibinfo {author} {\bibfnamefont {E.~A.}\ \bibnamefont {Gaffney}},
  \bibinfo {author} {\bibfnamefont {J.~R.}\ \bibnamefont {Blake}},\ and\
  \bibinfo {author} {\bibfnamefont {J.~C.}\ \bibnamefont {Kirkman-Brown}},\
  }\bibfield  {title} {\bibinfo {title} {Human sperm accumulation near
  surfaces: a simulation study},\ }\href
  {https://doi.org/10.1017/S0022112008004953} {\bibfield  {journal} {\bibinfo
  {journal} {J. Fluid Mech.}\ }\textbf {\bibinfo {volume} {621}},\ \bibinfo
  {pages} {289–320} (\bibinfo {year} {2009})}\BibitemShut {NoStop}%
\bibitem [{\citenamefont {Spagnolie}\ and\ \citenamefont
  {Lauga}(2012)}]{Spagnolie:2012}%
  \BibitemOpen
  \bibfield  {author} {\bibinfo {author} {\bibfnamefont {S.~E.}\ \bibnamefont
  {Spagnolie}}\ and\ \bibinfo {author} {\bibfnamefont {E.}~\bibnamefont
  {Lauga}},\ }\bibfield  {title} {\bibinfo {title} {{Hydrodynamics of
  self-propulsion near a boundary: predictions and accuracy of far-field
  approximations}},\ }\href {https://doi.org/10.1017/jfm.2012.101} {\bibfield
  {journal} {\bibinfo  {journal} {J. Fluid Mech.}\ }\textbf {\bibinfo {volume}
  {700}},\ \bibinfo {pages} {105} (\bibinfo {year} {2012})}\BibitemShut
  {NoStop}%
\bibitem [{\citenamefont {DiLuzio}\ \emph {et~al.}(2005)\citenamefont
  {DiLuzio}, \citenamefont {Turner}, \citenamefont {Mayer}, \citenamefont
  {Garstecki}, \citenamefont {Weibel}, \citenamefont {Berg},\ and\
  \citenamefont {Whitesides}}]{Diluzio:2005}%
  \BibitemOpen
  \bibfield  {author} {\bibinfo {author} {\bibfnamefont {W.~R.}\ \bibnamefont
  {DiLuzio}}, \bibinfo {author} {\bibfnamefont {L.}~\bibnamefont {Turner}},
  \bibinfo {author} {\bibfnamefont {M.}~\bibnamefont {Mayer}}, \bibinfo
  {author} {\bibfnamefont {P.}~\bibnamefont {Garstecki}}, \bibinfo {author}
  {\bibfnamefont {D.~B.}\ \bibnamefont {Weibel}}, \bibinfo {author}
  {\bibfnamefont {H.~C.}\ \bibnamefont {Berg}},\ and\ \bibinfo {author}
  {\bibfnamefont {G.~M.}\ \bibnamefont {Whitesides}},\ }\bibfield  {title}
  {\bibinfo {title} {Escherichia coli swim on the right-hand side},\ }\href
  {https://doi.org/10.1038/nature03660} {\bibfield  {journal} {\bibinfo
  {journal} {Nature}\ }\textbf {\bibinfo {volume} {435}},\ \bibinfo {pages}
  {1271} (\bibinfo {year} {2005})}\BibitemShut {NoStop}%
\bibitem [{\citenamefont {Lauga}\ \emph {et~al.}(2006)\citenamefont {Lauga},
  \citenamefont {DiLuzio}, \citenamefont {Whitesides},\ and\ \citenamefont
  {Stone}}]{Lauga:2006}%
  \BibitemOpen
  \bibfield  {author} {\bibinfo {author} {\bibfnamefont {E.}~\bibnamefont
  {Lauga}}, \bibinfo {author} {\bibfnamefont {W.~R.}\ \bibnamefont {DiLuzio}},
  \bibinfo {author} {\bibfnamefont {G.~M.}\ \bibnamefont {Whitesides}},\ and\
  \bibinfo {author} {\bibfnamefont {H.~A.}\ \bibnamefont {Stone}},\ }\bibfield
  {title} {\bibinfo {title} {{Swimming in circles: Motion of bacteria near
  solid boundaries}},\ }\href {https://doi.org/10.1529/biophysj.105.069401}
  {\bibfield  {journal} {\bibinfo  {journal} {Biophys. J.}\ }\textbf {\bibinfo
  {volume} {90}},\ \bibinfo {pages} {400} (\bibinfo {year} {2006})}\BibitemShut
  {NoStop}%
\bibitem [{\citenamefont {Utada}\ \emph {et~al.}(2014)\citenamefont {Utada},
  \citenamefont {Bennett}, \citenamefont {Fong}, \citenamefont {Gibiansky},
  \citenamefont {Yildiz}, \citenamefont {Golestanian},\ and\ \citenamefont
  {Wong}}]{Utada:2014}%
  \BibitemOpen
  \bibfield  {author} {\bibinfo {author} {\bibfnamefont {A.~S.}\ \bibnamefont
  {Utada}}, \bibinfo {author} {\bibfnamefont {R.~R.}\ \bibnamefont {Bennett}},
  \bibinfo {author} {\bibfnamefont {J.~C.}\ \bibnamefont {Fong}}, \bibinfo
  {author} {\bibfnamefont {M.~L.}\ \bibnamefont {Gibiansky}}, \bibinfo {author}
  {\bibfnamefont {F.~H.}\ \bibnamefont {Yildiz}}, \bibinfo {author}
  {\bibfnamefont {R.}~\bibnamefont {Golestanian}},\ and\ \bibinfo {author}
  {\bibfnamefont {G.~C.}\ \bibnamefont {Wong}},\ }\bibfield  {title} {\bibinfo
  {title} {Vibrio cholerae use pili and flagella synergistically to effect
  motility switching and conditional surface attachment},\ }\href
  {https://doi.org/10.1038/ncomms5913} {\bibfield  {journal} {\bibinfo
  {journal} {Nat. Commun.}\ }\textbf {\bibinfo {volume} {5}},\ \bibinfo {pages}
  {4913} (\bibinfo {year} {2014})}\BibitemShut {NoStop}%
\bibitem [{\citenamefont {Elgeti}\ \emph {et~al.}(2010)\citenamefont {Elgeti},
  \citenamefont {Kaupp},\ and\ \citenamefont {Gompper}}]{Elgeti:2010}%
  \BibitemOpen
  \bibfield  {author} {\bibinfo {author} {\bibfnamefont {J.}~\bibnamefont
  {Elgeti}}, \bibinfo {author} {\bibfnamefont {U.~B.}\ \bibnamefont {Kaupp}},\
  and\ \bibinfo {author} {\bibfnamefont {G.}~\bibnamefont {Gompper}},\
  }\bibfield  {title} {\bibinfo {title} {Hydrodynamics of sperm cells near
  surfaces},\ }\href@noop {} {\bibfield  {journal} {\bibinfo  {journal}
  {Biophys. J.}\ }\textbf {\bibinfo {volume} {99}},\ \bibinfo {pages} {1018}
  (\bibinfo {year} {2010})}\BibitemShut {NoStop}%
\bibitem [{\citenamefont {Di~Leonardo}\ \emph {et~al.}(2011)\citenamefont
  {Di~Leonardo}, \citenamefont {Dell'Arciprete}, \citenamefont {Angelani},\
  and\ \citenamefont {Iebba}}]{diLeonardo:2011}%
  \BibitemOpen
  \bibfield  {author} {\bibinfo {author} {\bibfnamefont {R.}~\bibnamefont
  {Di~Leonardo}}, \bibinfo {author} {\bibfnamefont {D.}~\bibnamefont
  {Dell'Arciprete}}, \bibinfo {author} {\bibfnamefont {L.}~\bibnamefont
  {Angelani}},\ and\ \bibinfo {author} {\bibfnamefont {V.}~\bibnamefont
  {Iebba}},\ }\bibfield  {title} {\bibinfo {title} {Swimming with an image},\
  }\href {https://doi.org/10.1103/PhysRevLett.106.038101} {\bibfield  {journal}
  {\bibinfo  {journal} {Phys. Rev. Lett.}\ }\textbf {\bibinfo {volume} {106}},\
  \bibinfo {pages} {038101} (\bibinfo {year} {2011})}\BibitemShut {NoStop}%
\bibitem [{\citenamefont {Lemelle}\ \emph {et~al.}(2013)\citenamefont
  {Lemelle}, \citenamefont {Palierne}, \citenamefont {Chatre}, \citenamefont
  {Vaillant},\ and\ \citenamefont {Place}}]{Lemelle:2013}%
  \BibitemOpen
  \bibfield  {author} {\bibinfo {author} {\bibfnamefont {L.}~\bibnamefont
  {Lemelle}}, \bibinfo {author} {\bibfnamefont {J.-F.}\ \bibnamefont
  {Palierne}}, \bibinfo {author} {\bibfnamefont {E.}~\bibnamefont {Chatre}},
  \bibinfo {author} {\bibfnamefont {C.}~\bibnamefont {Vaillant}},\ and\
  \bibinfo {author} {\bibfnamefont {C.}~\bibnamefont {Place}},\ }\bibfield
  {title} {\bibinfo {title} {Curvature reversal of the circular motion of
  swimming bacteria probes for slip at solid/liquid interfaces},\ }\href
  {https://doi.org/10.1039/C3SM51426A} {\bibfield  {journal} {\bibinfo
  {journal} {Soft Matter}\ }\textbf {\bibinfo {volume} {9}},\ \bibinfo {pages}
  {9759} (\bibinfo {year} {2013})}\BibitemShut {NoStop}%
\bibitem [{\citenamefont {Hu}\ \emph {et~al.}(2015{\natexlab{a}})\citenamefont
  {Hu}, \citenamefont {Wysocki}, \citenamefont {Winkler},\ and\ \citenamefont
  {Gompper}}]{Hu:2015}%
  \BibitemOpen
  \bibfield  {author} {\bibinfo {author} {\bibfnamefont {J.}~\bibnamefont
  {Hu}}, \bibinfo {author} {\bibfnamefont {A.}~\bibnamefont {Wysocki}},
  \bibinfo {author} {\bibfnamefont {R.~G.}\ \bibnamefont {Winkler}},\ and\
  \bibinfo {author} {\bibfnamefont {G.}~\bibnamefont {Gompper}},\ }\bibfield
  {title} {\bibinfo {title} {Physical sensing of surface properties by
  microswimmers – directing bacterial motion via wall slip},\ }\href
  {https://doi.org/10.1038/srep09586} {\bibfield  {journal} {\bibinfo
  {journal} {Sci. Rep.}\ }\textbf {\bibinfo {volume} {5}},\ \bibinfo {pages}
  {9586} (\bibinfo {year} {2015}{\natexlab{a}})}\BibitemShut {NoStop}%
\bibitem [{\citenamefont {Spagnolie}\ \emph {et~al.}(2015)\citenamefont
  {Spagnolie}, \citenamefont {Moreno-Flores}, \citenamefont {Bartolo},\ and\
  \citenamefont {Lauga}}]{Spagnolie:2015}%
  \BibitemOpen
  \bibfield  {author} {\bibinfo {author} {\bibfnamefont {S.~E.}\ \bibnamefont
  {Spagnolie}}, \bibinfo {author} {\bibfnamefont {G.~R.}\ \bibnamefont
  {Moreno-Flores}}, \bibinfo {author} {\bibfnamefont {D.}~\bibnamefont
  {Bartolo}},\ and\ \bibinfo {author} {\bibfnamefont {E.}~\bibnamefont
  {Lauga}},\ }\bibfield  {title} {\bibinfo {title} {Geometric capture and
  escape of a microswimmer colliding with an obstacle},\ }\href
  {https://doi.org/10.1039/C4SM02785J} {\bibfield  {journal} {\bibinfo
  {journal} {Soft Matter}\ }\textbf {\bibinfo {volume} {11}},\ \bibinfo {pages}
  {3396} (\bibinfo {year} {2015})}\BibitemShut {NoStop}%
\bibitem [{\citenamefont {Kuron}\ \emph {et~al.}(2019)\citenamefont {Kuron},
  \citenamefont {Stärk}, \citenamefont {Holm},\ and\ \citenamefont
  {de~Graaf}}]{Kuron:2019}%
  \BibitemOpen
  \bibfield  {author} {\bibinfo {author} {\bibfnamefont {M.}~\bibnamefont
  {Kuron}}, \bibinfo {author} {\bibfnamefont {P.}~\bibnamefont {Stärk}},
  \bibinfo {author} {\bibfnamefont {C.}~\bibnamefont {Holm}},\ and\ \bibinfo
  {author} {\bibfnamefont {J.}~\bibnamefont {de~Graaf}},\ }\bibfield  {title}
  {\bibinfo {title} {Hydrodynamic mobility reversal of squirmers near flat and
  curved surfaces},\ }\href {https://doi.org/10.1039/C9SM00692C} {\bibfield
  {journal} {\bibinfo  {journal} {Soft Matter}\ }\textbf {\bibinfo {volume}
  {15}},\ \bibinfo {pages} {5908} (\bibinfo {year} {2019})}\BibitemShut
  {NoStop}%
\bibitem [{\citenamefont {Takagi}\ \emph {et~al.}(2014)\citenamefont {Takagi},
  \citenamefont {Palacci}, \citenamefont {Braunschweig}, \citenamefont
  {Shelley},\ and\ \citenamefont {Zhang}}]{Takagi:2014}%
  \BibitemOpen
  \bibfield  {author} {\bibinfo {author} {\bibfnamefont {D.}~\bibnamefont
  {Takagi}}, \bibinfo {author} {\bibfnamefont {J.}~\bibnamefont {Palacci}},
  \bibinfo {author} {\bibfnamefont {A.~B.}\ \bibnamefont {Braunschweig}},
  \bibinfo {author} {\bibfnamefont {M.~J.}\ \bibnamefont {Shelley}},\ and\
  \bibinfo {author} {\bibfnamefont {J.}~\bibnamefont {Zhang}},\ }\bibfield
  {title} {\bibinfo {title} {Hydrodynamic capture of microswimmers into
  sphere-bound orbits},\ }\href {https://doi.org/10.1039/C3SM52815D} {\bibfield
   {journal} {\bibinfo  {journal} {Soft Matter}\ }\textbf {\bibinfo {volume}
  {10}},\ \bibinfo {pages} {1784} (\bibinfo {year} {2014})}\BibitemShut
  {NoStop}%
\bibitem [{\citenamefont {Rode}\ \emph {et~al.}(2019)\citenamefont {Rode},
  \citenamefont {Elgeti},\ and\ \citenamefont {Gompper}}]{Rode:2019}%
  \BibitemOpen
  \bibfield  {author} {\bibinfo {author} {\bibfnamefont {S.}~\bibnamefont
  {Rode}}, \bibinfo {author} {\bibfnamefont {J.}~\bibnamefont {Elgeti}},\ and\
  \bibinfo {author} {\bibfnamefont {G.}~\bibnamefont {Gompper}},\ }\bibfield
  {title} {\bibinfo {title} {Sperm motility in modulated microchannels},\
  }\href {https://doi.org/10.1088/1367-2630/aaf544} {\bibfield  {journal}
  {\bibinfo  {journal} {New J. Phys.}\ }\textbf {\bibinfo {volume} {21}},\
  \bibinfo {pages} {013016} (\bibinfo {year} {2019})}\BibitemShut {NoStop}%
\bibitem [{\citenamefont {Frangipane}\ \emph {et~al.}(2019)\citenamefont
  {Frangipane}, \citenamefont {Vizsnyiczai}, \citenamefont {Maggi},
  \citenamefont {Savo}, \citenamefont {Sciortino}, \citenamefont {Gigan},\ and\
  \citenamefont {Di~Leonardo}}]{Frangipane:2019}%
  \BibitemOpen
  \bibfield  {author} {\bibinfo {author} {\bibfnamefont {G.}~\bibnamefont
  {Frangipane}}, \bibinfo {author} {\bibfnamefont {G.}~\bibnamefont
  {Vizsnyiczai}}, \bibinfo {author} {\bibfnamefont {C.}~\bibnamefont {Maggi}},
  \bibinfo {author} {\bibfnamefont {R.}~\bibnamefont {Savo}}, \bibinfo {author}
  {\bibfnamefont {A.}~\bibnamefont {Sciortino}}, \bibinfo {author}
  {\bibfnamefont {S.}~\bibnamefont {Gigan}},\ and\ \bibinfo {author}
  {\bibfnamefont {R.}~\bibnamefont {Di~Leonardo}},\ }\bibfield  {title}
  {\bibinfo {title} {Invariance properties of bacterial random walks in complex
  structures},\ }\href {https://doi.org/10.1038/s41467-019-10455-y} {\bibfield
  {journal} {\bibinfo  {journal} {Nat. Commun.}\ }\textbf {\bibinfo {volume}
  {10}},\ \bibinfo {pages} {2442} (\bibinfo {year} {2019})}\BibitemShut
  {NoStop}%
\bibitem [{\citenamefont {Makarchuk}\ \emph {et~al.}(2019)\citenamefont
  {Makarchuk}, \citenamefont {Braz}, \citenamefont {Ara{\'{u}}jo},
  \citenamefont {Ciric},\ and\ \citenamefont {Volpe}}]{Makarchuk:2019}%
  \BibitemOpen
  \bibfield  {author} {\bibinfo {author} {\bibfnamefont {S.}~\bibnamefont
  {Makarchuk}}, \bibinfo {author} {\bibfnamefont {V.~C.}\ \bibnamefont {Braz}},
  \bibinfo {author} {\bibfnamefont {N.~A.~M.}\ \bibnamefont {Ara{\'{u}}jo}},
  \bibinfo {author} {\bibfnamefont {L.}~\bibnamefont {Ciric}},\ and\ \bibinfo
  {author} {\bibfnamefont {G.}~\bibnamefont {Volpe}},\ }\bibfield  {title}
  {\bibinfo {title} {{Enhanced propagation of motile bacteria on surfaces due
  to forward scattering}},\ }\href {https://doi.org/10.1038/s41467-019-12010-1}
  {\bibfield  {journal} {\bibinfo  {journal} {Nat. Commun.}\ }\textbf {\bibinfo
  {volume} {10}},\ \bibinfo {pages} {4110} (\bibinfo {year}
  {2019})}\BibitemShut {NoStop}%
\bibitem [{\citenamefont {Chepizhko}\ and\ \citenamefont
  {Franosch}(2019)}]{Chepizhko:2019}%
  \BibitemOpen
  \bibfield  {author} {\bibinfo {author} {\bibfnamefont {O.}~\bibnamefont
  {Chepizhko}}\ and\ \bibinfo {author} {\bibfnamefont {T.}~\bibnamefont
  {Franosch}},\ }\bibfield  {title} {\bibinfo {title} {Ideal circle
  microswimmers in crowded media},\ }\href {https://doi.org/10.1039/C8SM02030B}
  {\bibfield  {journal} {\bibinfo  {journal} {Soft Matter}\ }\textbf {\bibinfo
  {volume} {15}},\ \bibinfo {pages} {452} (\bibinfo {year} {2019})}\BibitemShut
  {NoStop}%
\bibitem [{\citenamefont {Stone}\ and\ \citenamefont
  {Samuel}(1996)}]{Stone:1996}%
  \BibitemOpen
  \bibfield  {author} {\bibinfo {author} {\bibfnamefont {H.~A.}\ \bibnamefont
  {Stone}}\ and\ \bibinfo {author} {\bibfnamefont {A.~D.~T.}\ \bibnamefont
  {Samuel}},\ }\bibfield  {title} {\bibinfo {title} {{Propulsion of
  Microorganisms by Surface Distortions}},\ }\href
  {https://doi.org/10.1103/PhysRevLett.77.4102} {\bibfield  {journal} {\bibinfo
   {journal} {Phys. Rev. Lett.}\ }\textbf {\bibinfo {volume} {77}},\ \bibinfo
  {pages} {4102} (\bibinfo {year} {1996})}\BibitemShut {NoStop}%
\bibitem [{\citenamefont {Masoud}\ and\ \citenamefont
  {Stone}(2019)}]{Masoud:2019}%
  \BibitemOpen
  \bibfield  {author} {\bibinfo {author} {\bibfnamefont {H.}~\bibnamefont
  {Masoud}}\ and\ \bibinfo {author} {\bibfnamefont {H.~A.}\ \bibnamefont
  {Stone}},\ }\bibfield  {title} {\bibinfo {title} {The reciprocal theorem in
  fluid dynamics and transport phenomena},\ }\href
  {https://doi.org/10.1017/jfm.2019.553} {\bibfield  {journal} {\bibinfo
  {journal} {J. Fluid Mech.}\ }\textbf {\bibinfo {volume} {879}},\ \bibinfo
  {pages} {P1} (\bibinfo {year} {2019})}\BibitemShut {NoStop}%
\bibitem [{\citenamefont {Kurzthaler}\ \emph {et~al.}(2020)\citenamefont
  {Kurzthaler}, \citenamefont {Zhu}, \citenamefont {Pahlavan},\ and\
  \citenamefont {Stone}}]{Kurzthaler:2020}%
  \BibitemOpen
  \bibfield  {author} {\bibinfo {author} {\bibfnamefont {C.}~\bibnamefont
  {Kurzthaler}}, \bibinfo {author} {\bibfnamefont {L.}~\bibnamefont {Zhu}},
  \bibinfo {author} {\bibfnamefont {A.~A.}\ \bibnamefont {Pahlavan}},\ and\
  \bibinfo {author} {\bibfnamefont {H.~A.}\ \bibnamefont {Stone}},\ }\bibfield
  {title} {\bibinfo {title} {Particle motion nearby rough surfaces},\ }\href
  {https://doi.org/10.1103/PhysRevFluids.5.082101} {\bibfield  {journal}
  {\bibinfo  {journal} {Phys. Rev. Fluids}\ }\textbf {\bibinfo {volume} {5}},\
  \bibinfo {pages} {082101} (\bibinfo {year} {2020})}\BibitemShut {NoStop}%
\bibitem [{\citenamefont {Kamrin}\ \emph {et~al.}(2010)\citenamefont {Kamrin},
  \citenamefont {Bazant},\ and\ \citenamefont {Stone}}]{Kamrin:2010}%
  \BibitemOpen
  \bibfield  {author} {\bibinfo {author} {\bibfnamefont {K.}~\bibnamefont
  {Kamrin}}, \bibinfo {author} {\bibfnamefont {M.~Z.}\ \bibnamefont {Bazant}},\
  and\ \bibinfo {author} {\bibfnamefont {H.~A.}\ \bibnamefont {Stone}},\
  }\bibfield  {title} {\bibinfo {title} {{Effective slip boundary conditions
  for arbitrary periodic surfaces: The surface mobility tensor}},\ }\href
  {https://doi.org/10.1017/S0022112010001801} {\bibfield  {journal} {\bibinfo
  {journal} {J. Fluid Mech.}\ }\textbf {\bibinfo {volume} {658}},\ \bibinfo
  {pages} {409} (\bibinfo {year} {2010})}\BibitemShut {NoStop}%
\bibitem [{\citenamefont {Lighthill}(1952)}]{Lighthill:1952}%
  \BibitemOpen
  \bibfield  {author} {\bibinfo {author} {\bibfnamefont {M.~J.}\ \bibnamefont
  {Lighthill}},\ }\bibfield  {title} {\bibinfo {title} {On the squirming motion
  of nearly spherical deformable bodies through liquids at very small reynolds
  numbers},\ }\href {https://doi.org/10.1002/cpa.3160050201} {\bibfield
  {journal} {\bibinfo  {journal} {Commun. Pure Appl. Math.}\ }\textbf {\bibinfo
  {volume} {5}},\ \bibinfo {pages} {109} (\bibinfo {year} {1952})}\BibitemShut
  {NoStop}%
\bibitem [{\citenamefont {Shaik}\ and\ \citenamefont
  {Ardekani}(2017)}]{Shaik:2017}%
  \BibitemOpen
  \bibfield  {author} {\bibinfo {author} {\bibfnamefont {V.~A.}\ \bibnamefont
  {Shaik}}\ and\ \bibinfo {author} {\bibfnamefont {A.~M.}\ \bibnamefont
  {Ardekani}},\ }\bibfield  {title} {\bibinfo {title} {Motion of a model
  swimmer near a weakly deforming interface},\ }\href
  {https://doi.org/10.1017/jfm.2017.285} {\bibfield  {journal} {\bibinfo
  {journal} {J. of Fluid Mech.}\ }\textbf {\bibinfo {volume} {824}},\ \bibinfo
  {pages} {42–73} (\bibinfo {year} {2017})}\BibitemShut {NoStop}%
\bibitem [{\citenamefont {Ishikawa}\ \emph {et~al.}(2006)\citenamefont
  {Ishikawa}, \citenamefont {Simmonds},\ and\ \citenamefont
  {Pedley}}]{Ishikawa:2006}%
  \BibitemOpen
  \bibfield  {author} {\bibinfo {author} {\bibfnamefont {T.}~\bibnamefont
  {Ishikawa}}, \bibinfo {author} {\bibfnamefont {M.~P.}\ \bibnamefont
  {Simmonds}},\ and\ \bibinfo {author} {\bibfnamefont {T.~J.}\ \bibnamefont
  {Pedley}},\ }\bibfield  {title} {\bibinfo {title} {Hydrodynamic interaction
  of two swimming model micro-organisms},\ }\href
  {https://doi.org/10.1017/S0022112006002631} {\bibfield  {journal} {\bibinfo
  {journal} {J. Fluid Mech.}\ }\textbf {\bibinfo {volume} {568}},\ \bibinfo
  {pages} {119–160} (\bibinfo {year} {2006})}\BibitemShut {NoStop}%
\bibitem [{\citenamefont {Blake}\ and\ \citenamefont
  {Chwang}(1974)}]{Blake:1974}%
  \BibitemOpen
  \bibfield  {author} {\bibinfo {author} {\bibfnamefont {J.~R.}\ \bibnamefont
  {Blake}}\ and\ \bibinfo {author} {\bibfnamefont {A.~T.}\ \bibnamefont
  {Chwang}},\ }\bibfield  {title} {\bibinfo {title} {{Fundamental singularities
  of viscous flow}},\ }\href {https://doi.org/10.1007/bf02353701} {\bibfield
  {journal} {\bibinfo  {journal} {J. Eng. Math}\ }\textbf {\bibinfo {volume}
  {8}},\ \bibinfo {pages} {23} (\bibinfo {year} {1974})}\BibitemShut {NoStop}%
\bibitem [{\citenamefont {Kim}\ and\ \citenamefont {Karrila}(2005)}]{Kim:2005}%
  \BibitemOpen
  \bibfield  {author} {\bibinfo {author} {\bibfnamefont {S.}~\bibnamefont
  {Kim}}\ and\ \bibinfo {author} {\bibfnamefont {S.}~\bibnamefont {Karrila}},\
  }\href {https://books.google.com/books?id=\_8llnUUGo0wC} {\emph {\bibinfo
  {title} {Microhydrodynamics: Principles and Selected Applications}}},\
  Butterworth - Heinemann Series in Chemical Engineering\ (\bibinfo
  {publisher} {Dover Publications},\ \bibinfo {year} {2005})\BibitemShut
  {NoStop}%
\bibitem [{\citenamefont {Hu}\ \emph {et~al.}(2015{\natexlab{b}})\citenamefont
  {Hu}, \citenamefont {Yang}, \citenamefont {Gompper},\ and\ \citenamefont
  {Winkler}}]{Hu:2015:SM}%
  \BibitemOpen
  \bibfield  {author} {\bibinfo {author} {\bibfnamefont {J.}~\bibnamefont
  {Hu}}, \bibinfo {author} {\bibfnamefont {M.}~\bibnamefont {Yang}}, \bibinfo
  {author} {\bibfnamefont {G.}~\bibnamefont {Gompper}},\ and\ \bibinfo {author}
  {\bibfnamefont {R.~G.}\ \bibnamefont {Winkler}},\ }\bibfield  {title}
  {\bibinfo {title} {Modelling the mechanics and hydrodynamics of swimming e.
  coli},\ }\href@noop {} {\bibfield  {journal} {\bibinfo  {journal} {Soft
  {M}atter}\ }\textbf {\bibinfo {volume} {11}},\ \bibinfo {pages} {7867}
  (\bibinfo {year} {2015}{\natexlab{b}})}\BibitemShut {NoStop}%
\bibitem [{\citenamefont {Secchi}\ \emph {et~al.}(2020)\citenamefont {Secchi},
  \citenamefont {Vitale}, \citenamefont {Mi{\~n}o}, \citenamefont {Kantsler},
  \citenamefont {Eberl}, \citenamefont {Rusconi},\ and\ \citenamefont
  {Stocker}}]{Secchi:2020}%
  \BibitemOpen
  \bibfield  {author} {\bibinfo {author} {\bibfnamefont {E.}~\bibnamefont
  {Secchi}}, \bibinfo {author} {\bibfnamefont {A.}~\bibnamefont {Vitale}},
  \bibinfo {author} {\bibfnamefont {G.~L.}\ \bibnamefont {Mi{\~n}o}}, \bibinfo
  {author} {\bibfnamefont {V.}~\bibnamefont {Kantsler}}, \bibinfo {author}
  {\bibfnamefont {L.}~\bibnamefont {Eberl}}, \bibinfo {author} {\bibfnamefont
  {R.}~\bibnamefont {Rusconi}},\ and\ \bibinfo {author} {\bibfnamefont
  {R.}~\bibnamefont {Stocker}},\ }\bibfield  {title} {\bibinfo {title} {The
  effect of flow on swimming bacteria controls the initial colonization of
  curved surfaces},\ }\href@noop {} {\bibfield  {journal} {\bibinfo  {journal}
  {Nat. Commun.}\ }\textbf {\bibinfo {volume} {11}},\ \bibinfo {pages} {1}
  (\bibinfo {year} {2020})}\BibitemShut {NoStop}%
\end{thebibliography}%

\end{document}